\documentclass[fleqn,usenatbib]{mnras}
\usepackage{newtxtext, newtxmath, xcolor}
\usepackage{array}

\usepackage[T1]{fontenc}

\DeclareRobustCommand{\VAN}[3]{#2}
\let\VANthebibliography\thebibliography
\def\thebibliography{\DeclareRobustCommand{\VAN}[3]{##3}\VANthebibliography}

\usepackage{graphicx}	      % Including figure files
\usepackage{amsmath}	      % Advanced maths commands
\usepackage{amssymb}	      % Extra maths symbols
\usepackage{hyperref}         % Insert clickable urls
\usepackage[normalem]{ulem}   % For \sout{}
\usepackage{eucal}            % Replace \mathcal with Euler script version
\usepackage{tabularx}
\usepackage{multirow}
\usepackage{anyfontsize}

\newcommand{\parenth}[1]{\left(#1\right)}
\newcommand{\bracket}[1]{\left[#1\right]}
\newcommand{\Msun}{\mathrm{M}_\odot}
\newcommand{\percent}{ per cent}
\defcitealias{Yang+2012}{Y12}
\defcitealias{Tempel+2017}{T17}
\defcitealias{Rodriguez_Merchan2020}{R\&M20}
\defcitealias{Coil+2017}{C17}
\defcitealias{Berti+2019}{B19}
\defcitealias{Berti+2021}{B21}

\title[Clustering Statistics and the Galaxy--Halo Connection]{Star-forming and Quiescent Central Galaxies Cluster Similarly: Implications for the Galaxy--Halo Connection}

\author[James Kakos et al.]{James Kakos,$^{1}$\thanks{E-mail: jpkakos@gmail.com}
Aldo Rodr\'iguez-Puebla,$^{2}$\thanks{E-mail: apuebla@astro.unam.mx}
Joel R. Primack,$^{1}$
Sandra M. Faber,$^{3}$
David C. Koo,$^{3}$
\newauthor{Peter Behroozi,$^{4}$
and Vladimir Avila-Reese$^{2}$}
\\
$^{1}$Physics Department, University of California, Santa Cruz, CA 95060, USA\\
$^{2}$Universidad Nacional Aut\'onoma de M\'exico, Instituto de Astronom\'ia, A. P. 70-264, 04510, CDMX, M\'exico\\
$^{3}$UCO/Lick Observatory, Department of Astronomy and Astrophysics, University of California, Santa Cruz, CA 95064, USA\\
$^{4}$ Department of Astronomy and Steward Observatory, University of Arizona, Tucson, AZ 85721, USA
}
\date{}

\pubyear{2024}

\begin{document}
\label{firstpage}
\pagerange{\pageref{firstpage}--\pageref{lastpage}}
\maketitle

% Abstract of the paper, in 250 words or less
\begin{abstract}
We measure the clustering of low-redshift SDSS galaxies as a function of stellar mass ($10.0<\log(M_*/\Msun)<11.5$) and specific star formation rate (sSFR) and compare the results to models of the galaxy--halo connection. We find that the auto-correlation functions of central galaxies exhibit little dependence on sSFR, with the well-known stronger clustering of quiescent galaxies mainly attributable to satellites. Because halo assembly history is known to affect distinct halo clustering, this result implies that there is little net correlation between halo assembly history and central galaxy sSFR. However, cross-correlations with satellites are stronger for quiescent centrals than star-forming centrals, consistent with quiescent centrals having more satellites in their haloes at fixed $M_*$, as found in SDSS group catalogues. We model the galaxy--halo connection in an $N$-body simulation by assigning sSFRs to central galaxies in three different ways. Two of the models depend on halo assembly history (being based on halo accretion rate or concentration), while the third is independent of halo assembly history (being based on peak halo circular velocity, $V_\text{peak}$, a proxy for halo mass). All three models replicate the observed auto-correlations of central galaxies, while only the $V_\text{peak}$ model reproduces the observed cross-correlations with satellites. This further suggests that the effects of halo assembly history may not be easily seen in auto-correlations of centrals and implies that a more complete understanding of central galaxy clustering may require more than auto-correlations of centrals alone. Additionally, the good agreement with the $V_\text{peak}$ model supports the idea that quiescent galaxies reside in more massive haloes than star-forming galaxies at fixed $M_*$.
\end{abstract}

%%%%%%%%%%%%%%%%%%%%%%%%%%%%%%%%%%%%%%%%%%%%%%%%%%

%%%%%%%%%%%%%%%%% BODY OF PAPER %%%%%%%%%%%%%%%%%%

\section{Introduction} \label{section:introduction}
According to the $\Lambda$CDM paradigm, galaxy formation and evolution take place within massive dark matter haloes. A key component of our understanding of galaxy formation and evolution within this context is the stellar-to-halo mass relation (SHMR). The SHMR relates the mass of stars within a galaxy to the mass of its host halo. This serves as a vital tool in connecting the observed properties of galaxies with the underlying dark matter haloes -- known as the galaxy--halo connection -- allowing us to explore the processes governing galaxy formation, the impact of feedback mechanisms, galaxy bias, and the cosmological context in which galaxies form and evolve \citep[for a review, see][]{Somerville_Dave2015}.

The SHMR is derived by measuring the stellar masses of a large sample of galaxies and associating them with their corresponding halo masses, which are inferred through various methods such as gravitational lensing \citep{Mandelbaum+2006, Mandelbaum+2016}, galaxy clustering \citep{Berlind_Weinberg2002}, galaxy kinematics \citep{More+2011, WojtakMamon2013, Lange+2019}, empirical modeling \citep{Behroozi+2019}, abundance matching techniques \citep{Conroy+2006}, or any combination of these methods \citep[for a review, see][]{Wechsler_Tinker2018}. These probes of the galaxy--halo connection provide a statistical description of the SHMR, which is assumed to be an increasing one-to-one monotonic relationship. In addition, the SHMR exhibits scatter, indicating that there is a range of stellar masses for a given halo mass. This scatter may arise from a combination of factors such as the stochasticity of the star formation process, varying merger histories, and the influence of environmental effects on galaxy formation. Previous attempts to constrain the scatter around the SHMR have found it to be of the order $\sim$0.15~dex \citep[see, e.g.,][]{Rodriguez-Puebla+2015, Behroozi+2019, Porras-Valverde+2023}.

Whatever is influencing a galaxy's position in the SHMR should be related to the assembly history of the galaxy and, ultimately, its star formation activity (or color), i.e., linked to its position within the specific star formation rate (sSFR)--stellar mass plane. Observationally, several previous studies have identified a robust segregation in halo mass at fixed stellar mass, where quiescent/red central galaxies inhabit more massive haloes compared to star-forming/blue central galaxies \citep[e.g.,][]{More+2011, Tinker+2013, Rodriguez-Puebla+2015, Mandelbaum+2016, Lange+2019}. At least part of the explanation for this segregation is the fact that while a galaxy's star formation may cease, its host halo can continue to grow hierarchically, especially in the case of more massive haloes. Consequently, for a given stellar mass, galaxies that ceased star formation earlier tend to have larger halo masses, even if they had high star formation efficiency during their active star-forming phase. Notably, this segregation in the SHMR becomes more pronounced in massive galaxies, with quiescent central galaxies residing in haloes that are a factor of $\sim$2 more massive than haloes of star-forming central galaxies \citep{More+2011, Rodriguez-Puebla+2015, Mandelbaum+2016}.

Another way to describe the SHMR segregation is that at fixed halo mass, $M_\text{h}$, star-forming galaxies have a higher stellar mass, $M_{*}$, than quiescent galaxies. Previous studies by \citet*{Moster+2018, Moster+2020} have found an opposite relation, that quiescent galaxies have a \emph{higher} stellar mass than star-forming galaxies at a fixed halo mass. This would seem to imply that quiescent galaxies reside in haloes of \emph{lower} mass than star-forming galaxies at a fixed stellar mass. However, these authors point out that due to the scatter in the SHMR, $\langle M_*(M_\text{h})\rangle$ cannot simply be inverted to obtain $\langle M_\text{h}(M_*)\rangle$ (see also \citealt*{Rodriguez-Puebla+2013}; \citealt{Rodriguez-Puebla+2015}). The combined effect of Eddington bias and a higher fraction of star-forming galaxies at lower masses results in the average halo mass being higher for quenched galaxies than star-forming galaxies at a fixed stellar mass, which is consistent with observational constraints. This phenomenon has been described in the literature as the inversion problem \citep{Cui+2021}. In addition to an opposite relation, we note that some papers have found little evidence of such a segregation in the SHMR \citep[see, e.g., fig.~38 of][]{Behroozi+2019}.

Do specific halo properties determine a galaxy's position in the SHMR, or is it instead influenced by other stochastic factors? Is the strong segregation observed in the SHMR a genuine phenomenon? If so, does it align with the notion of halo assembly bias -- the concept that the clustering behavior of haloes depends not only on halo mass but also on formation history -- as a main driver of the assembly histories of galaxies? Two-point correlation functions are well-explored tools for investigating these questions. By assuming that the centers of haloes and sub-haloes serve as the locations of central and satellite galaxies, we can focus on modeling how galaxies inhabit haloes of varying masses to understand galaxy clustering. In particular, if the segregation is indeed present, it could be interpreted that the SHMR is primarily a reflection of halo mass as the determining factor for stellar mass and sSFR. This would suggest that, at a given stellar mass, the enhanced clustering of quiescent galaxies could be attributed to their occupancy of more massive haloes \citep{Rodriguez-Puebla+2015}. On the other hand, if differences in clustering are influenced by other halo properties, two possibilities emerge:
\begin{enumerate}
    \item The scatter around the SHMR may be merely random variation unrelated to galaxy assembly history, with halo assembly history being the key factor determining how galaxies cluster \citep[e.g.,][]{Hearin_Watson_2013}.
    \item There could exist a segregation in the SHMR, and clustering information is shared between this segregation and halo assembly bias.
\end{enumerate}
These considerations present scenarios that can be tested using various approaches for measuring two-point correlation functions, as discussed in the main body of this paper.

It is observationally established that red/quiescent galaxies tend to be more clustered than blue/star-forming galaxies \citep[see, e.g.,][]{Li+2006, Heinis+2009, Zehavi+2011, Coil+2017, Berti+2021}. Recent works by \citet{Coil+2017} and \citet{Berti+2021} studied the auto-correlation functions (ACFs) of PRIMUS, DEEP2, and SDSS galaxies as a function of their sSFR and stellar mass. They found that in bins of stellar mass, the ACF decreases monotonically as sSFR increases, including within the star-forming and the quiescent populations, which \citet{Berti+2021} refer to as intrasequence relative bias. They interpret these differences in galaxy clustering as a function of sSFR as evidence that the scatter in galaxy sSFR is physically connected to the large-scale cosmic density field.

The degree to which halo properties are responsible for the differing clustering of star-forming and quiescent galaxies is still uncertain \citep{Blanton+2007, Tinker+2011, ODonnell+2021, ODonnell+2022}. The simplest models of the galaxy--halo connection depend solely on halo mass to determine the properties of galaxies \citep[for a discussion see][]{Wechsler_Tinker2018}. However, there has been an effort to expand models of the galaxy--halo connection to include halo properties beyond mass \citep[e.g.,][]{Hearin_Watson_2013, Masaki+2013, Becker+2015}. Some such studies aiming to identify halo properties that influence the regulation of SFRs in galaxies have highlighted two potentially key factors: halo accretion rate and halo concentration.

Halo accretion rate controls the influx of gas into a galaxy's interstellar medium \citep[see, e.g.,][]{Avila-Reese+2000, Bouche+2010, DekelMandelker2014, WetzelNagai2015, Rodriguez-Puebla+2016a}. One might expect halo accretion rate to be linked to SFR, as having more gas available may allow a galaxy to form more stars. It has also been shown that, at a fixed halo mass, haloes with lower accretion rates tend to be in higher density environments at low redshifts \citep{Lee+2017}, so it may also be expected that quiescent galaxies are more clustered than star-forming galaxies when modeling sSFR based on halo accretion rate. This, indeed, has been shown by \citet{Becker+2015} where the relative clustering amplitudes of quiescent and star-forming galaxies agree with observations. Additionally, the dispersion in the halo accretion rate has been shown to reproduce the width of the SFR distribution in star-forming galaxies at various redshifts \citep{DekelMandelker2014, Rodriguez-Puebla+2016a}.

Halo concentration, on the other hand, is linked to the timing of gas infall into the halo, with more concentrated haloes experiencing earlier infall (see, e.g., \citealt*{Avila-Reese+1998}; \citealt{Wechsler+2002, Gao+2004}; \citealt*{Dutton+2010a}; \citealt{Matthee+2019}). A prevailing explanation for the difference in clustering for quiescent and star-forming galaxies is that galaxies that form earlier reach the ends of their life cycles (i.e., become quiescent) earlier, and earlier-formed haloes tend to be more clustered. At a fixed halo mass, haloes with high concentration tend to be the ones that formed earlier and are more clustered \citep{Wechsler+2006, Gao+2008, Montero-Dorta+2021}, so it is natural to connect halo concentration to galaxy SFR. Indeed, models based on halo concentration have successfully reproduced the observed clustering of red and blue galaxies \citep{Hearin_Watson_2013, Masaki+2013}.

In both cases of halo accretion rate and halo concentration, halo assembly bias is assumed to be important for understanding the distribution and behavior of galaxies within dark matter haloes \citep{ShethTormen2004, GaoSpringelWhite2005, Wechsler+2006, Montero-Dorta+2021}. It is important to note that existing descriptions of assembly bias and density fields have been primarily developed for distinct haloes (i.e., excluding sub-haloes). This would correspond to a sample of only central galaxies, excluding satellites. Despite sharing the same host halo, centrals and satellites are expected to have distinct formation and evolutionary histories, which makes the connection between satellites and the haloes in which they reside more complex. We highlight the fact that past studies of correlation function trends with sSFR or galaxy color have generally not made distinctions between centrals and satellites. The findings of \citet{Coil+2017} were obtained using samples of all galaxies. As a follow-up, \citet{Berti+2019} studied clustering trends of PRIMUS galaxies using isolated primary (IP) galaxies as proxies for centrals. While they do also find that quiescent IPs are more clustered than star-forming IPs, competing biases from satellite contamination and environmental incompleteness leave some uncertainty as to the extent of the clustering difference for centrals. Similar to \citet{Coil+2017}, \citet{Berti+2021} used samples of all galaxies to study clustering trends of SDSS galaxies. However, they showed that a modified version of the \textsc{UniverseMachine} model \citep{Behroozi+2019} agrees with their ACF results and that in this model, central galaxies contribute substantially to the dependence of clustering on sSFR at a given stellar mass. In this paper, we will test the sSFR-dependence on galaxy clustering, focusing on central galaxies only. This will allow for a more direct comparison with the theoretical framework of the galaxy--halo connection. As we will show, the presence of satellite galaxies introduces additional variability in the correlations. Furthermore, we find that ACFs alone may not be sufficient to discriminate models of the galaxy--halo connection when considering only central galaxies, while cross-correlation functions (CCFs) between central and satellite galaxies prove to be a powerful tool in this regard.

This paper is organized as follows. In Section~\ref{section:data}, we describe the main data set we use, including the cuts that were made to define our samples and the group catalogues used to identify central and satellite galaxies. In Section~\ref{section:methods}, we describe our methods for calculating correlation functions and binning the data into sub-samples as a function of stellar mass and sSFR. Section~\ref{section:results} presents our results of ACFs and CCFs within these various sub-samples. Using these results, we test different models of the galaxy--halo connection in Section~\ref{section:galaxy_halo_connection}. In Section~\ref{section:discussion}, we summarize our results and provide a discussion of their robustness and comparisons with the literature. Finally, in Section~\ref{section:conclusions}, we present our conclusions.

\section{Data} \label{section:data}
\subsection{Sloan Digital Sky Survey}
In this study, we use observations from the Sloan Digital Sky Survey (SDSS) with galaxy redshifts taken from SDSS Data Release 7 \citep{Abazajian+2009}. Stellar masses and star formation rates (SFRs) are taken from the MPA-JHU catalogue \citep{Kauffmann+2003, Brinchmann+2004} and have a \citet{Kroupa+2001} initial mass function. \citet{Brinchmann+2004} used observed emission lines, including H$\alphaup$, within the central fiber of the SDSS and modeled them based on the \citet{Charlot_Longhetti_2001} stellar population synthesis model. They showed that using the standard \citet{Kennicutt1998rev} conversion factor from H$\alphaup$ to SFR is a good average correction for most of the star-forming galaxies. The same is not true for quiescent galaxies, however, for which their SFRs are mostly given by the D4000 break.

\subsection{Group catalogues} \label{section:group_catalogs}
One of the main goals of this paper is to quantify, separately, the contributions of central and satellite galaxies to the observed two-point auto-correlation function (ACF) as well as their cross-correlation function (CCF). Therefore, an important designation we use for our sample is whether a galaxy is a central or satellite. In this paper, we utilize the halo-based group catalogue by \citet[hereafter \citetalias{Yang+2012}]{Yang+2012} as our primary database to identify central and satellite galaxies within the SDSS. It is important to note that we exclusively employ their spectroscopic group sample, comprising a total of 593,227 galaxies at $z\leq0.2$. Since galaxy group finders operate in redshift space, they are inherently susceptible to errors from redshift-space distortions, making it difficult to achieve a perfect galaxy-to-group assignment and introducing systematic uncertainties in the identifications. To ensure the robustness of our results, we also incorporate two additional group catalogues: \citet[hereafter \citetalias{Tempel+2017}]{Tempel+2017} and \citet[hereafter \citetalias{Rodriguez_Merchan2020}]{Rodriguez_Merchan2020}, containing 571,291 and 648,480 galaxies, respectively, at $z\le0.2$. It is essential to clarify that our primary aim in using these alternative catalogues is not to critique or evaluate which group finder performs best in identifying groups but rather to examine whether all three catalogues yield consistent results. In each of these catalogues, we define central galaxies as the most massive galaxy within a group in terms of stellar mass, with any remaining galaxies in the group taken to be satellites. Note that by this definition, isolated galaxies that have no satellites will be considered centrals.

\subsection{Data selection} \label{section:data_selection}
To calculate two-point correlation functions, we use a random catalogue of the SDSS provided by \citet{Blanton+2005}. The random catalogue has 1 million points distributed with constant surface density over the area of SDSS. The SDSS projected sky distribution has irregular edges and some holes where no objects are catalogued. In order to ensure consistency between the observations and the random catalogue when calculating two-point correlation functions, we trim the edges of both distributions following \citet{Varela+2012} and \citet[see also \citealp{Dragomir+2018}]{Cebrian+2014}. The following four cuts are applied:
\begin{enumerate}
    \item Southern limit: $\delta > 0^\circ$
    \item Western limit: $\delta > -2.555556(\alpha-131^\circ)$
    \item Eastern limit: $\delta > 1.70909(\alpha-235^\circ)$
    \item Northern limit: $\delta < \arcsin\left[ \frac{0.93232\sin(\alpha-95\overset{\circ}{.}9)}{\sqrt{1-\left[ 0.93232\cos(\alpha-95\overset{\circ}{.}9)\right] ^2}} \right]$.
\end{enumerate}
These cuts can be seen in fig.~2 of \citet{Varela+2012} and fig.~1 of \citet{Cebrian+2014}. Additionally, we mask out the holes in both the observations and the random catalogue with small rectangular cuts over the affected regions.

After applying these cuts to the data, we then narrow our selection down to the five mass bins. Each bin will correspond to a volume-limited sample that is complete in stellar mass over the range of the bin. The mass bins cover the range $10.0 < \log(M_*/\Msun) < 11.5$, where the two lower mass bins have widths of 0.375~dex and the three higher mass bins have widths of 0.25~dex (as listed in Table~\ref{tab:subvolume_stats}). We choose the lower mass bins to be wider to improve statistics since the lower masses require smaller volumes to maintain completeness.

\begin{figure*}
    \centering
    \includegraphics[width=\linewidth]{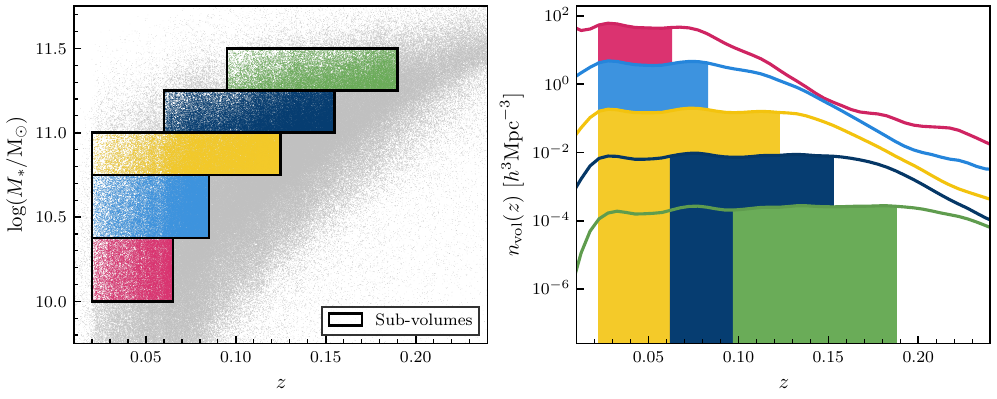}
    \caption{\textit{Left panel}: stellar mass as a function of redshift for the SDSS. The rectangles with colored points show the five volume-limited samples used for measuring correlation functions. \textit{Right panel}: number densities as a function of redshift for the five sub-volumes. The solid lines show the distributions over the entire redshift range, while the colored regions show the selected redshift ranges spanned by the sub-volumes. The redshift limits of the sub-volumes are determined by traveling in each direction from the maximum of the distribution to points where the counts drop to 50\percent{} of the maximum. This process was applied to various sub-samples within each sub-volume to ensure completeness in all of them (see Section~\ref{section:data_selection} and Fig.~\ref{fig:completeness_subsamples} for details). Note that each successive distribution from front to back is scaled up by an additional order of magnitude for clarity.}
    \label{fig:completeness}
\end{figure*}

To define the sub-volumes, we follow the `Volume1' procedure in Appendix~A of \citetalias{Yang+2012}.\footnote{Additionally, \citetalias{Yang+2012} defines a `Volume2' procedure following \citet{vandenBosch+2008}, which accounts for the differing selection effects of red and blue galaxies in flux-limited surveys, owing to their different mass-to-light ratios. We find our resulting volumes to be generally consistent with this alternative completeness limit.} The redshift limits of the sub-volumes are defined based on a minimum threshold in the galaxy number density, $n_\text{vol}(z)$, for a given sub-volume. In this way, sub-volumes will have relatively flat $n_\text{vol}(z)$ distributions, i.e., number densities that are near-constant. To determine the redshift limits, in each mass bin, we locate the redshift corresponding to the peak of $n_\text{vol}(z)$ and then travel to lower and higher redshifts until the distribution falls to 50\percent{} of the maximum. For this process, we calculate $n_\text{vol}(z)$ in redshift bins of width $dz=0.005$ and apply a Gaussian smoothing with $\sigma=dz$ to the overall distribution. This smoothing helps to avoid spikes within any single redshift bin of $n_\text{vol}(z)$ biasing the results of the process (e.g., regions of low density due to cosmic variance, rather than a lack of observations, can cross the 50\percent{} threshold resulting in an artificially smaller sub-volume). Using a threshold of 50\percent{} was chosen as a balance between completeness and statistics. Choosing a higher threshold would yield sub-volumes with a higher degree of completeness; however, the two-point statistics would become less reliable as the sub-volumes would be smaller with fewer galaxies. Within each sub-volume, this process is repeated for the individual sub-samples of galaxies that we work with: all, central, satellite, star-forming, green valley, and quiescent galaxies. We choose the redshift limits that allow us to satisfy our completeness condition in all sub-samples simultaneously within a mass bin. 

The five sub-volumes are shown in the left panel of Fig.~\ref{fig:completeness} along with the number densities in the right panel. The colored lines show the number densities across the full redshift range in each mass bin, and the shaded regions show the limits that define the sub-volumes based on our completeness condition. Note that the number densities are scaled up by increasing orders of magnitude from front to back for clarity. A more detailed look at the number densities for the individual sub-samples can be seen in Fig.~\ref{fig:completeness_subsamples}. The result of our selection process is a catalogue of 208,730 galaxies across the five sub-volumes. Specifics for each sub-volume are described in Table~\ref{tab:subvolume_stats}.

\begin{table}
    \centering
    \caption{Comparison of the five sub-volumes used. Stellar mass bins cover the range $10.0 < \log(M_*/\mathrm{M}_\odot) < 11.5$, with the lower two mass bins having widths of 0.375~dex and the upper three mass bins having widths of 0.25~dex. The redshift ranges correspond to the ranges where the number densities of galaxies are at least 50\percent{} of the maximum in a given mass bin (see Section~\ref{section:data_selection} and Fig.~\ref{fig:completeness_subsamples} for details).}
    \begin{tabular}{c c c}
        \hline
        \rule{0pt}{8pt}$M_*$ Limits $\bracket{\log(M_*/\Msun)}$ & $z$ Limits & $N_\text{gal}$\\
        \hline
        \rule{0pt}{8pt}\phantom{00}10.0 -- 10.375 & \phantom{0}0.02 -- 0.065 & 23,969\\
        10.375 -- 10.75\phantom{0} & \phantom{0}0.02 -- 0.085 & 48,221\\
        \phantom{0}10.75 -- 11.0\phantom{00} & \phantom{0}0.02 -- 0.125 & 57,533\\
        \phantom{00}11.0 -- 11.25\phantom{0} & \phantom{0}0.06 -- 0.155 & 51,738\\
        \phantom{0}11.25 -- 11.5\phantom{00} & 0.095 -- 0.19\phantom{0} & 27,269\\
        \hline
    \end{tabular}
    \label{tab:subvolume_stats}
\end{table}

\begin{figure}
    \centering
    \includegraphics[width=\linewidth]{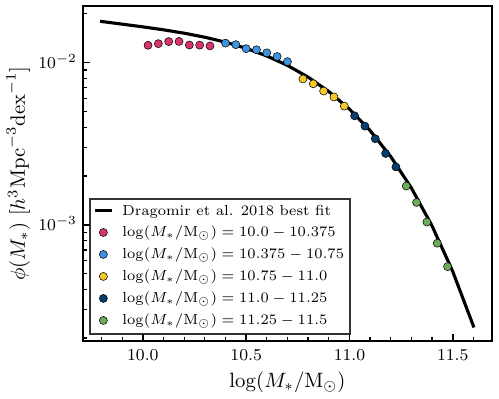}
    \caption{Galaxy stellar mass function (GSMF) of our five sub-volumes compared to a double Schechter function with the best-fitting parameters from \citet{Dragomir+2018}. Our sub-volumes match well with the SDSS GSMF, with a slight discrepancy at the lowest masses. This is the result of a trade-off between completeness and sufficient statistics to calculate correlation functions.}
    \label{fig:gsmf}
\end{figure}

As a check of the completeness in our resulting sub-volumes, we calculate the galaxy stellar mass function (GSMF) across our full mass range. Fig.~\ref{fig:gsmf} shows the GSMF in each of our five sub-volumes (colored points) compared to a double Schechter function using the \citet{Dragomir+2018} best-fitting parameters of SDSS (black line). Overall, we find generally good agreement between our sub-volumes and the fit, though there is some discrepancy at low mass. This is likely the result of the 50\percent{} limit being too low and creating a volume that is too large for fainter low-mass galaxies. While this leads to a lower level of completeness than our other sub-volumes, the larger volume is required to obtain reliable two-point statistics.

\section{Methods} \label{section:methods}
\subsection{Clustering measures}
Galaxy clustering was measured using two-point correlation functions, which measure the excess probability over random of finding pairs of galaxies separated by a distance $r$. To mitigate the effects of redshift-space distortions, the correlation function is calculated in a projected area broken into components perpendicular to $(r_\text{p})$ and parallel to $(r_\piup)$ the line of sight. Assuming two galaxies are positioned at $\mathbf{s_1}$ and $\mathbf{s_2}$, a line-of-sight vector can be defined as $\mathbf{l} = \frac12(\mathbf{s_1}+\mathbf{s_2})$ and a separation vector can be defined as $\mathbf{s}=\mathbf{s_1}-\mathbf{s_2}$. With these definitions, $r_\text{p}$ and $r_\piup$ can be calculated as
\begin{align}
    &r_\text{p} = \sqrt{\mathbf{s}\cdot\mathbf{s}-r_\piup^2}\\
    &r_\piup = \frac{\mathbf{s}\cdot\mathbf{l}}{\lvert\mathbf{l}\rvert}.
\end{align}
Correlation functions are calculated using the \citet{Landy_Szalay1993} estimator
\begin{equation}
    \xi(r_\text{p}, r_\piup) = \frac{\text{D}_1\text{D}_2 - \text{D}_1\text{R}_2 - \text{D}_2\text{R}_1 + \text{R}_1\text{R}_2}{\text{R}_1\text{R}_2}
    \label{eq:LS_estimator}
\end{equation}
where D$_1$D$_2$ is the data--data pair counts, D$_1$R$_2$ and D$_2$R$_1$ are the data--random pair counts, and R$_1$R$_2$ is the random-random pair counts. For CCFs, the subscripts denote the two different data sets that are being cross-correlated. For ACFs, the subscripts are all the same and equation~(\ref{eq:LS_estimator}) simplifies to
\begin{equation}
    \xi(r_\text{p}, r_\piup) = \frac{\text{DD} - \text{2DR} + \text{RR}}{\text{RR}}.
\end{equation}
These are then integrated over $r_\piup$ to find the two-dimensional projected correlation functions
\begin{equation}
    w_\text{p}(r_\text{p}) = 2\int_0^{r_\piup^\text{max}} \xi(r_\text{p}, r_\piup)\mathrm{d}r_\piup.
\end{equation}
For our calculations, we integrate to $r_\piup^\text{max} = 20~h^{-1}$Mpc. The calculations were done using \textsc{corrfunc} \citep{Corrfunc2020} in 10 $r_\text{p}$ bins from $0.1~h^{-1}$Mpc to $20~h^{-1}$Mpc. Error bars for correlation functions show the 1$\sigma$ error after running 200 bootstrap samples. Since we will show results of both projected ACFs and CCFs, we will use $w_\text{a}$ to denote projected ACFs and $w_\text{c}$ to denote projected CCFs.

Correlation functions can be affected on small scales by fibre collisions. In the SDSS, galaxies within 55~arcsec of each other cannot receive fibres on the same plate, but some regions (roughly a third of the sky) were tiled with overlapping plates. We tested for the impact of fibre collisions by upweighting pair counts for galaxies that had pairs on small scales by a factor of 3. The largest impact was, as expected, on scales below $r_\text{p}\approx 0.1~\text{Mpc}$ for galaxies around $\log(M_*/\Msun) = 10$ and scales below $r_\text{p}\approx 0.2~\text{Mpc}$ for galaxies around $\log(M_*/\Msun) = 11$, with changes at the few-percent level on larger scales. Hence, we do not expect fibre collisions to affect conclusions for any of the correlation function analyses in this paper.

\subsection{Fitting the star-forming main sequence}
In this section, we describe how we define the star-forming main sequence (SFMS). We follow \citet[][see also \citealp{Rodriguez-Puebla+2020} and \citealp{Fang+2018}]{Stephenson+2024} and briefly describe the method below. This definition will be the basis for how we break down the sSFR--$M_*$ plane into a grid of sub-samples (see Section~\ref{section:grid}). The idea is to create bins in sSFR based on distance (in dex) from the mean of the SFMS at a fixed $M_*$. To do this, we use an iterative process of fitting the sSFR--$M_*$ relation of star-forming galaxies until the fit becomes stable. The general process is as follows:
\begin{enumerate}
    \item Fit a straight line to the data sample of $\log(\text{sSFR/yr}^{-1})$ vs $\log(M_*/\Msun)$.
    \item Shift the line down 0.45~dex in $\log(\text{sSFR/yr}^{-1})$ and select all galaxies above the line to be the new sample.
    \item Repeat until the fit parameters are stable to a maximum tolerance of $10^{-3}$.
\end{enumerate}
Once stable parameters are obtained for the straight line fit, we calculate the median sSFR of all galaxies above the line (i.e., star-forming galaxies) as a function of $M_*$ in 10 mass bins. These medians are used to fit a function of the form
\begin{equation}
    \langle\log[\text{sSFR}_\text{MS}(M_*)]\rangle = \log\parenth{\frac{\psi_0}{M_*}} - \log\bracket{1+\parenth{\frac{M_*}{M_0}}^{\gamma}}
    \label{eq:sfms_mean}
\end{equation}
where $\psi_0$, $M_0$, and $\gamma$ are fitting parameters \citep{Lee+2015}. For our data set, we find the best-fitting parameters $\log(\psi_0/\Msun\text{yr}^{-1})=0.829$, $\log(M_0/\Msun)=10.914$, and $\gamma=-0.897$. Equation~(\ref{eq:sfms_mean}) defines the mean of the SFMS, which we use to create all other sSFR bins. We take all galaxies 0.45~dex below this curve or higher to be the SFMS.We note that because this definition is derived by fitting across our five sub-volumes that span different redshift ranges, it inherently captures any redshift dependence in the SFMS.

\subsection{The sSFR--\texorpdfstring{$M_*$}{M*} grid} \label{section:grid}
\begin{figure}
    \centering
    \includegraphics[width=\linewidth]{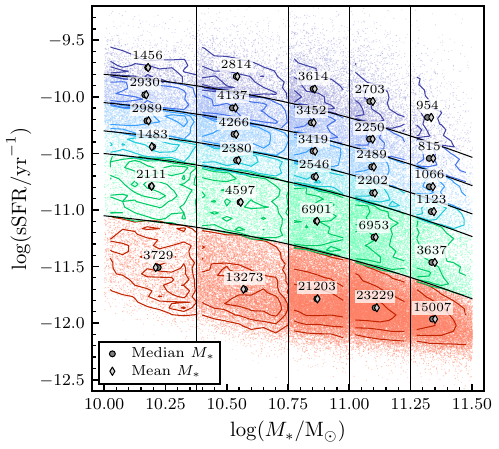}
    \caption{Scatter plot of the sSFR--$M_*$ plane for \citetalias{Yang+2012} central galaxies showing the grid binning scheme. Correlation functions are calculated within individual cells or combinations of cells in the grid. The data are broken horizontally into the five sub-volumes shown in Fig.~\ref{fig:completeness} and vertically by distance from the SFMS (specifics shown in Table~\ref{tab:delta_ms_bins}). The text boxes show the number of galaxies within each grid cell. Gray circles and diamonds show the median and mean stellar masses and median sSFR within each cell. Overall, we find the median and mean stellar masses to be similar in value and close to constant as a function of sSFR, which helps remove mass effects in the trends of the correlation functions. See Table~\ref{tab:sample_statistics} for details on galaxy counts and mean and median values of stellar mass and $\Delta$MS for both centrals and satellites.}
    \label{fig:ssfr-Mstar_plane}
\end{figure}

In order to systematically study ACFs and CCFs in the sSFR--$M_*$ plane, we break our galaxy sample into a grid binning by stellar mass and distance from the mean of the SFMS. The five mass bins defined in Section~\ref{section:data_selection} are used to mitigate the effects of mass dependence on the correlation functions and isolate the dependence on sSFR. Fig.~\ref{fig:ssfr-Mstar_plane} shows the mean and median masses within each cell of the grid. In every grid cell, the mean and median mass are similar in value to each other, and in any given mass bin, they each remain relatively constant as a function of sSFR (the largest variation within any mass bin is $\lesssim$~0.05 dex). This ensures that, within a given mass bin, any disparities in the two-point correlation functions across the sSFR ranges will be \emph{solely attributable} to their distinct clustering properties rather than being influenced by internal mass effects within the bin.

Using equation~(\ref{eq:sfms_mean}) as an initial bin edge, we choose four additional parallel curves to break sSFR into six bins. In this way, we are binning galaxies by their distance from the mean of the SFMS, which we define as $\Delta$MS, where $\Delta\text{MS} \equiv\ \log[\text{sSFR}(M_*)] - \langle\log[\text{sSFR}_\text{MS}(M_*)]\rangle$. We follow the bin definitions used in \citet{Stephenson+2024} and describe them here. The divisions for star-forming galaxies are made based on the assumption that the SFMS has a width of $\sigma\sim0.25$~dex (see, e.g., \citealp{Speagle+2014}). Highly star-forming (HSF) galaxies lie at least 1$\sigma$ above the mean of the SFMS and potentially include starburst galaxies. Upper main sequence (UMS) and lower main sequence (LMS) galaxies lie above and below the mean of the SFMS, respectively, but are constrained to be within $\pm1\sigma$. Bottom of the main sequence (BMS) galaxies lie at least 1$\sigma$ below the mean of the SFMS but above -0.45 dex. Green valley (GV) galaxies are those between 0.45 dex and 1 dex below the mean of the SFMS, and quiescent (Q) galaxies are all galaxies more than 1 dex below the mean of the SFMS. These bin definitions are summarized in Table~\ref{tab:delta_ms_bins}. Fig.~\ref{fig:ssfr-Mstar_plane} shows the sSFR--$M_*$ plane with our grid plotted as black lines on top of it. The second horizontal curve from the top shows our fit of the mean of the SFMS, from which all other horizontal curves are derived. See Table~\ref{tab:sample_statistics} for details on galaxy counts and mean and median values of stellar mass and $\Delta$MS for both centrals and satellites in the grid.

\section{Results} \label{section:results}
The primary objective of this paper is to investigate the clustering properties of galaxies across the sSFR--$M_*$ plane, as shown in Fig.~\ref{fig:ssfr-Mstar_plane}. Previous studies have primarily focused on the clustering properties of galaxies by categorizing them into star-forming/blue or quiescent/red galaxies \citep[see, e.g.,][]{Li+2006, Zehavi+2011, Guo+2011, Hearin_Watson_2013, Tinker+2013}, with only a few exceptions examining correlations across the sSFR--$M_*$ plane \citep{Coil+2017, Berti+2021}. To the best of our knowledge, there have been no reports on two-point correlations across the sSFR--$M_*$ plane when distinguishing between central and satellite galaxies or on the cross-correlations between centrals and satellites. As we will discuss in this paper, a detailed examination of cross-correlations is essential for understanding how galaxies are connected to their dark matter haloes. In the following sections, we present and discuss our results on galaxy clustering.

As mentioned in Section~\ref{section:group_catalogs}, we use three different group catalogues for this work. All results in Section~\ref{section:results} were calculated using each of the group catalogues, and while we do not show all of these results, we find that they all produce clustering trends that are consistent with each other and lead to the same conclusions we draw in this paper. A discussion comparing the group catalogues can be found in Section~\ref{section:group_catalog_comparison}.

Finally, note that for clarity in all correlation function figures, we do not plot points where the error exceeds the value of the point itself (unless specified otherwise). Additionally, the label at the top of each column in each correlation function figure shows the value of the center of that mass bin in log space as a general guide for the stellar masses to which each column corresponds.

\begin{table}
    \centering
    \caption{List of sub-samples in the sSFR--$M_*$ plane and the corresponding $\Delta$MS bin definitions.}
    \begin{tabular}{l c}
        \hline
        \rule{0pt}{8pt}Sub-sample & $\Delta$MS Bin [dex]\\
        \hline
        \rule{0pt}{8pt}Highly star-forming (HSF) & \phantom{-0.00<0}$\Delta$MS > 0.25\phantom{-}\\
        Upper main sequence (UMS) & \phantom{-0.0}0 < $\Delta$MS < 0.25\phantom{-}\\
        Lower main sequence (LMS) & -0.25 < $\Delta$MS < 0\phantom{-0.0}\\
        Bottom of the main sequence (BMS) & -0.45 < $\Delta$MS < -0.25\\
        Green valley (GV) & \phantom{.00}-1 < $\Delta$MS < -0.45\\
        Quiescent (Q) & \phantom{-0.00<0}$\Delta$MS < -1\phantom{.00}\\
        \hline
    \end{tabular}
    \label{tab:delta_ms_bins}
\end{table}

\subsection{Auto-correlation functions} \label{section:acf}
\begin{figure*}
    \centering
    \includegraphics[width=\linewidth]{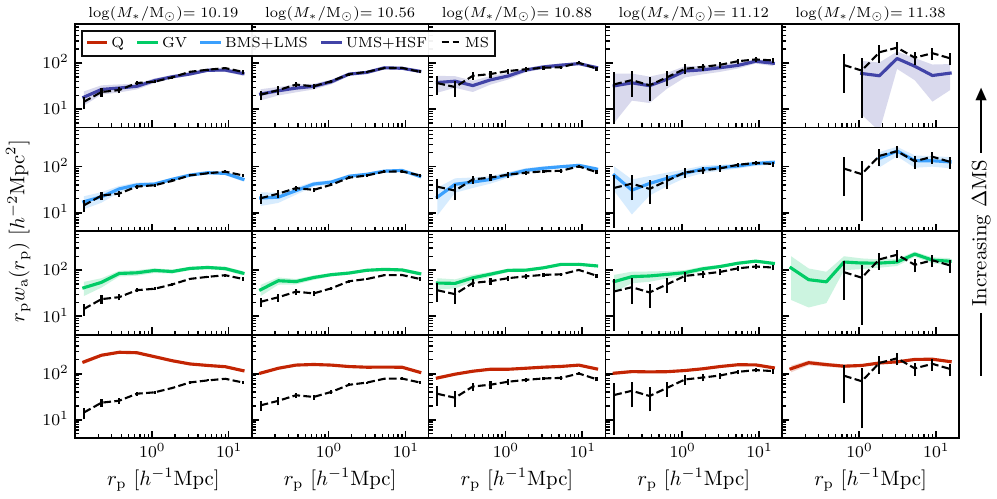}
    \caption{Projected auto-correlation functions (ACFs) of all galaxies as a function of $\Delta$MS and $M_*$. The rows correspond from top to bottom to UMS+HSF galaxies (dark blue), BMS+LMS galaxies (light blue), GV galaxies (green), and quiescent galaxies (red). The colored lines show the results within a grid cell, and the black dashed lines show the ACFs of all galaxies in the center of the SFMS (LMS+UMS). While showing no evolution across the SFMS, there is a clear trend of increasing clustering amplitude as $\Delta$MS decreases below the SFMS. The magnitude of this effect tends to decrease with mass due to the differing galaxy selections in the different sub-volumes. Greater clustering at small separations is primarily the result of the presence of satellite galaxies, and the relative fraction of satellite galaxies within each mass bin decreases as the mass increases (see Fig.~\ref{fig:all_ACF_components}).}
    \label{fig:all_ACFs}
\end{figure*}
\begin{figure*}
    \centering
    \includegraphics[width=\linewidth]{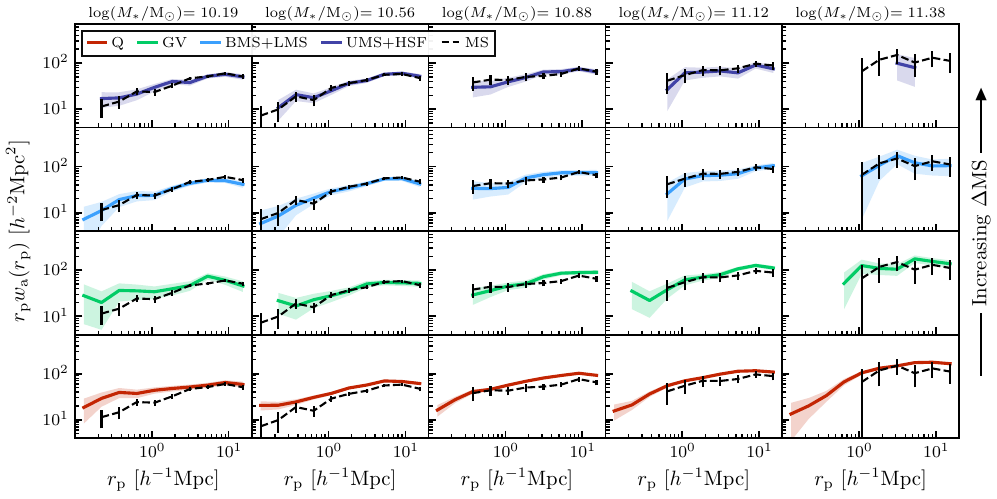}
    \caption{Projected auto-correlation functions (ACFs) of central galaxies as a function of $\Delta$MS and $M_*$. The rows correspond from top to bottom to UMS+HSF galaxies (dark blue), BMS+LMS galaxies (light blue), GV galaxies (green), and quiescent galaxies (red). The colored lines show the results within a grid cell, and the black dashed lines show the ACFs of central galaxies in the center of the SFMS (LMS+UMS). We observe a drastic reduction in the trends as a function of $\Delta$MS compared to the ACFs of all galaxies (see Fig.~\ref{fig:all_ACFs}). The remaining trend is only clear in the most quiescent galaxies. However, because these are ACFs of central galaxies, only scales larger than $r_\text{p}\sim 1~h^{-1}$Mpc should be considered as multiple centrals cannot exist within the same halo; signals below this scale can be caused by the projection used to calculate $w_\text{a}$ and by satellites misidentified as centrals. We note that when using \citetalias{Tempel+2017}, which we believe contains a more pure sample of only centrals, small-scale signals are generally removed (see Fig.~\ref{fig:tempel_centrals_ACFs}). This is discussed further in Section~\ref{section:group_catalog_comparison}.}
    \label{fig:centrals_ACFs}
\end{figure*}
\begin{figure*}
    \centering
    \includegraphics[width=\linewidth]{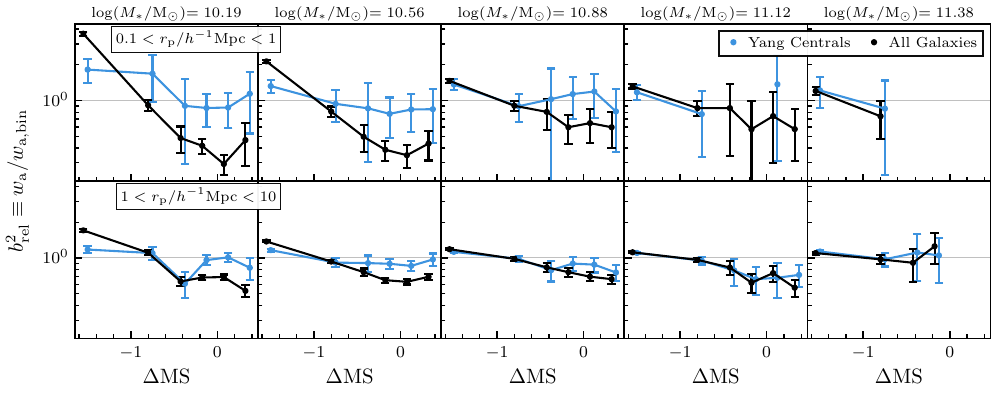}
    \caption{Relative bias as a function of $\Delta$MS for \citetalias{Yang+2012} centrals (blue) and all galaxies (black). We define the bias of any sub-sample relative to the auto-correlation function of the entire mass bin collectively. As such, only the \emph{change} in bias should be compared across mass bins. The top row shows the bias averaged over small scales ($0.1 < r_\text{p} / h^{-1}\text{Mpc} < 1$) and the bottom row shows the bias averaged over large scales ($1 < r_\text{p} / h^{-1}\text{Mpc} < 10$). In both cases, we see a reduction in the change in bias as a function of $\Delta$MS for centrals compared to all galaxies. Note that low-mass, low-$\Delta$MS signals at small scales in the \citetalias{Yang+2012} centrals may be caused by satellite contamination. See Section~\ref{section:group_catalog_comparison} and Fig.~\ref{fig:tempel_centrals_ACFs} for more details on this.}
    \label{fig:centrals_ACFs_bias}
\end{figure*}
\begin{figure*}
    \centering
    \includegraphics[width=\linewidth]{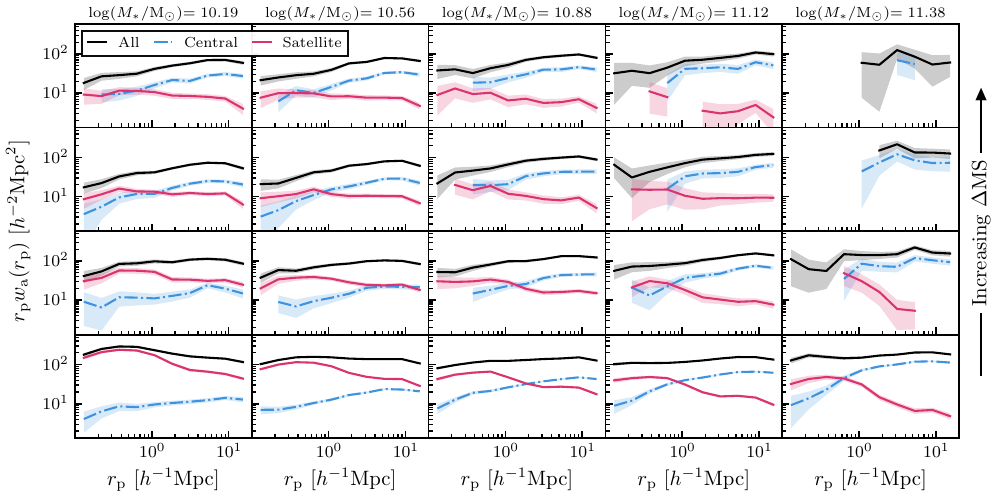}
    \caption{Projected auto-correlation functions (ACFs) of all (black), central (dash-dot blue), and satellite (magenta) galaxies as a function of $\Delta$MS and $M_*$. ACFs of centrals and satellites are scaled by their fractional numbers according to equation~(\ref{eq:weighted_cf}) to show their relative contributions to the overall ACF. The rows correspond from top to bottom to UMS+HSF galaxies, BMS+LMS galaxies, GV galaxies, and quiescent galaxies. The increased clustering with decreasing $\Delta$MS in the ACFs of all galaxies is driven primarily by satellite galaxies, which are highly clustered and tend to be quiescent. On the other hand, the ACFs of central galaxies show little to no trend with $\Delta$MS at a fixed stellar mass.}
    \label{fig:all_ACF_components}
\end{figure*}

Fig.~\ref{fig:all_ACFs} shows the ACFs of all galaxies (centrals and satellites together) as a function of $\Delta$MS. The dark and light blue lines in the upper two rows correspond to galaxies in the upper (HSF+UMS) and lower (LMS+BMS) halves of the SFMS, respectively. The green lines in the third row correspond to green valley galaxies, and the red lines in the bottom row correspond to quiescent galaxies. In each column, there is a black dashed line that remains the same in every row. These lines represent the ACFs of galaxies in the center of the SFMS (LMS+UMS) and serve as a reference to see how the ACFs evolve from high to low $\Delta$MS at a fixed mass. We observe a strong and consistent trend of increasing clustering going below the SFMS. This effect is most pronounced at lower masses where the clustering amplitude increases up to an order of magnitude above the SFMS amplitude. In every mass bin, the increase in the clustering amplitude occurs most strongly at smaller separations, i.e., in the one-halo term where the highly clustered nature of satellites has the greatest impact on the ACFs. This signal, however, becomes weaker in the higher mass bins as there are fewer satellites in the sub-samples at higher masses.

We now investigate the relative contributions of central and satellite galaxies to this observed trend in the ACFs. Fig.~\ref{fig:centrals_ACFs} shows the ACFs of central galaxies only. With this sample, we notice a drastic reduction in the trends as a function of $\Delta$MS. The remaining weak trend is only somewhat noticeable in the most quiescent centrals, especially at lower masses. Fig.~\ref{fig:centrals_ACFs_bias} shows the relative bias as a function of $\Delta$MS for \citetalias{Yang+2012} centrals (blue) compared to all galaxies (black). The relative bias is calculated as
\begin{equation}
    b_\text{rel} = \sqrt{\frac{w_\text{a}}{w_\text{a,bin}}},
    \label{eq:bias}
\end{equation}
where $w_\text{a}$ is the ACF of a sub-sample and $w_\text{a,bin}$ is the ACF of the full sample in a given mass bin. Since $w_\text{a,bin}$ is different for each mass bin, only the \emph{change} in bias should be compared across mass bins. The top row shows the bias averaged over small scales ($0.1 < r_\text{p} / h^{-1}\text{Mpc} < 1$) and the bottom shows the bias averaged over large scales ($1 < r_\text{p} / h^{-1}\text{Mpc} < 10$). We see that the change in bias as a function of $\Delta$MS is reduced for centrals compared to all galaxies at both small and large scales (for specific values of $b_\text{rel}$, see Table~\ref{tab:bias}). Note that since these ACFs are of centrals only, we focus primarily on scales larger than $r_\text{p} \sim 1~h^{-1}$Mpc as multiple centrals cannot occupy the same halo. Signals below this scale are likely the result of the projection used to calculate the ACFs and potentially satellites that are misidentified as centrals. When using \citetalias{Tempel+2017}, which we believe contains a more pure sample of only centrals, the small-scale signals in the ACFs are generally removed (see Fig.~\ref{fig:tempel_centrals_ACFs}). This is discussed further in Section~\ref{section:group_catalog_comparison}. 

Focusing on the bottom row of Fig.~\ref{fig:centrals_ACFs_bias}, we find that almost every central galaxy sub-sample (blue points) is consistent with $b^2_\text{rel}=1$, showing there is little significant change in the ACFs with $\Delta$MS. The only sub-samples that are more than 3$\sigma$ away from $b^2_\text{rel}=1$ correspond to the quiescent samples in the central three mass bins: $\log (M_*/\Msun) = 10.56$, 10.88 and 11.12. Despite these samples being the most significantly deviated from $b^2_\text{rel}=1$, the values themselves remain quite close to 1 ($1.08\pm0.02$, $1.06\pm0.01$, and $1.05\pm0.01$). On the other hand, we can compare these results to the biases of all galaxies (black points). For this, we focus primarily on the two lowest mass bins because the satellite fractions are the highest at lower masses, thus creating the most difference between the central galaxy and all galaxy samples. At higher masses, the black points converge toward the blue points as the samples become more similar. In the two lowest mass bins, we find that the quiescent samples have biases $\gtrsim$10$\sigma$ above $b^2_\text{rel}=1$ ($1.31\pm0.03$ and $1.17\pm0.01$) and all star-forming samples are at least 4$\sigma$ below $b^2_\text{rel}=1$, with values generally around $b^2_\text{rel}\sim0.8$ and uncertainties ranging from $\sigma=0.02$--0.03. Similar results are obtained when using the \citetalias{Tempel+2017} and \citetalias{Rodriguez_Merchan2020} group catalogues.

We conclude that the strong trend and main contributor of increased clustering amplitude below the SFMS is therefore the product of satellite galaxies in the sample. Further, the ratio of satellite galaxies to central galaxies increases with distance below the SFMS. This effect is strongest at lower masses, and these galaxy samples are precisely where we see the strongest clustering in Fig.~\ref{fig:all_ACFs}.

Fig.~\ref{fig:all_ACF_components} shows the weighted ACFs of all (black), central (dash-dot blue), and satellite (magenta) galaxies to illustrate the relative contributions of centrals and satellites to the overall ACF of all galaxies. Formally, this is given by
\begin{equation}
    w^{\phantom{}}_{\text{a,\,all}} = f^{2}_{\vphantom{l}\text{cen}} w^{\phantom{}}_{\vphantom{l}\text{a,\,cen}} + f^{2}_{\vphantom{l}\text{sat}} w^{\phantom{}}_{\vphantom{l}\text{a,\,sat}} + 2f^{\phantom{}}_{\vphantom{l}\text{cen}} f^{\phantom{}}_{\vphantom{l}\text{sat}} w^{\phantom{}}_{\vphantom{l}\text{c,\,{cen--sat}}}
    \label{eq:weighted_cf}
\end{equation}
where $f_{\text{cen}}$ and $f_{\text{sat}}$ are the fractions of central and satellite galaxies, respectively.\footnote{To derive equation~(\ref{eq:weighted_cf}), we use the fact that for two samples $i$ and $j$, the number of galaxy pairs between them, $N_{ij}$, can be related to the number density of objects in each sample, $n_i$ and $n_j$, and the correlation function between them, $w_{ij}$, as $N_{ij} = n_in_j(1+w_{ij})$. Note that for auto-correlations, $i=j$, which yields pair counts $N_{i} = \frac12 n_i^2(1+w_{i})$ where the factor of $\frac12$ is to avoid double counting. Equation~(\ref{eq:weighted_cf}) can then be derived using the relations $N_\text{all} = N_{\text{cen}} + N_{\text{sat}} + N_{\text{cen-sat}}$ and $n_\text{all} = n_\text{cent} + n_\text{sat}$ \citep[see also][]{Zehavi+2011}.} The last term in equation~(\ref{eq:weighted_cf}) is a weighted CCF between centrals and satellites. We have plotted separately the first two terms but omitted this last term in Fig.~\ref{fig:all_ACF_components} to ensure clarity regarding the relative contributions of the ACFs. We will revisit central--satellite CCFs in Section~\ref{section:ccf}.

In Fig.~\ref{fig:all_ACF_components}, the ACFs of centrals (dash-dot blue lines) are seen to remain relatively constant as a function of $\Delta$MS while the ACFs of satellites (magenta lines) grow substantially with decreasing $\Delta$MS, particularly at small separations and most pronounced at lower masses where most satellites reside. This reaffirms our conclusion that satellites are the main driver of the trends shown in Fig.~\ref{fig:all_ACFs} and explains why the trend weakens with increasing mass. As we consider bins of higher mass, the satellite fraction drops, causing the ACFs to become more weighted by contributions from centrals, which generally do not change with $\Delta$MS.

These results are consistent with the idea that satellites are the main drivers of environmental effects \citet[][see also \citealp{Peng+2012, Kovac+2014}]{Peng+2010}. \citet{Kovac+2014}, e.g., found that the red fraction of centrals is produced almost entirely by mass-quenching processes, with minimal dependence on environmental effects. The red fraction of satellites, however, required additional environmental quenching to explain observations. If we use the ACFs as proxies for density, the minimally enhanced clustering of quiescent centrals over star-forming centrals and the strong dependence of satellite clustering on sSFR are both consistent with the findings of \cite{Peng+2010, Peng+2012} and \citet{Kovac+2014}.

In the existing literature, there is substantial empirical evidence indicating that, on average, a galaxy's stellar mass increases monotonically with halo mass. Because halo clustering also increases with halo mass -- often referred to as halo bias -- one would expect that the clustering of central galaxies should similarly increase with stellar mass. Indeed, this is what we observe in Figs~\ref{fig:centrals_ACFs} and \ref{fig:all_ACF_components}. Apart from halo bias, the clustering amplitude of haloes can vary based on other factors, such as formation redshift and accretion history, a phenomenon known as assembly bias (see, e.g., \citealp{Wechsler+2006}). The empirical evidence presented in Figs~\ref{fig:centrals_ACFs} and \ref{fig:all_ACF_components} -- that is, the almost nonexistent dependence of the central galaxy ACFs with $\Delta$MS at a fixed stellar mass -- suggests that assembly bias is not producing any significant net effect in central galaxy clustering in the sSFR--$M_*$ plane.

\subsection{Cross-correlation functions} \label{section:ccf}
\begin{figure*}
    \centering
    \includegraphics[width=\linewidth]{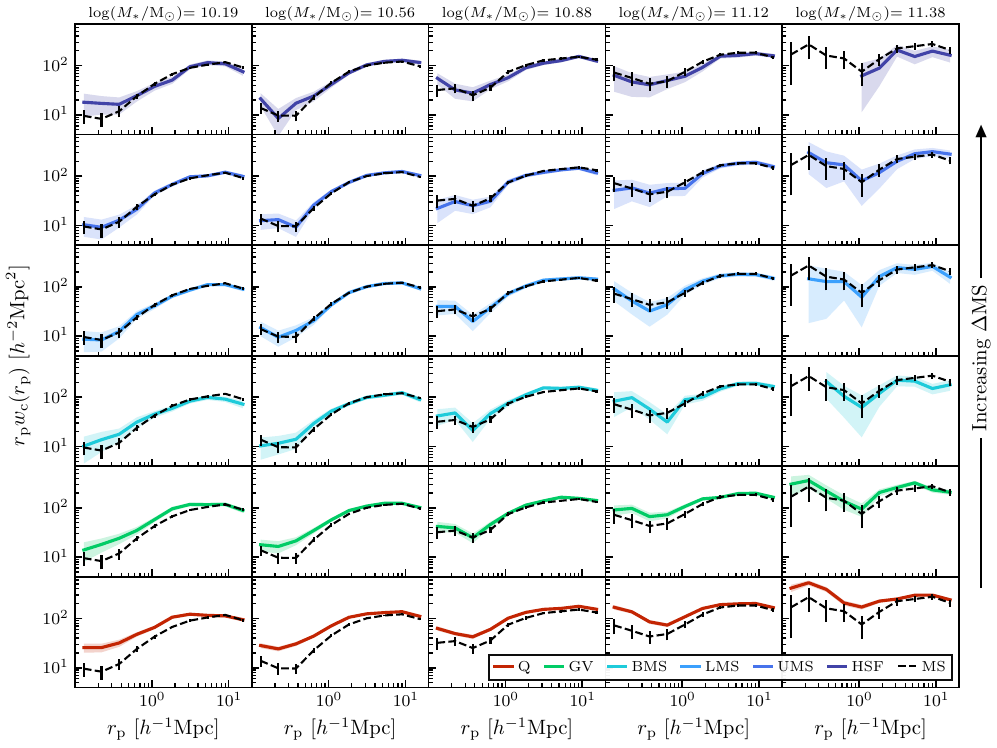}
    \caption{Projected cross-correlation functions (CCFs) of central galaxies as a function of $\Delta$MS and $M_*$. The centrals of a given cell in the grid are cross-correlated with all other satellites in the same mass bin. For comparison, a black dashed line is included that represents the CCF of all centrals in the center of the SFMS (LMS+UMS) with all satellites in the same mass bin. While there is no strong trend in the CCFs across the SFMS (top four rows in blue), we do see a trend of increasing clustering at small separations for GV centrals (green) and a further increase for quiescent centrals (red). Because this increase occurs at small separations, it suggests that more quiescent centrals will tend to have more satellite galaxies.}
    \label{fig:CCFs}
\end{figure*}
\begin{figure*}
    \centering
    \includegraphics[width=\linewidth]{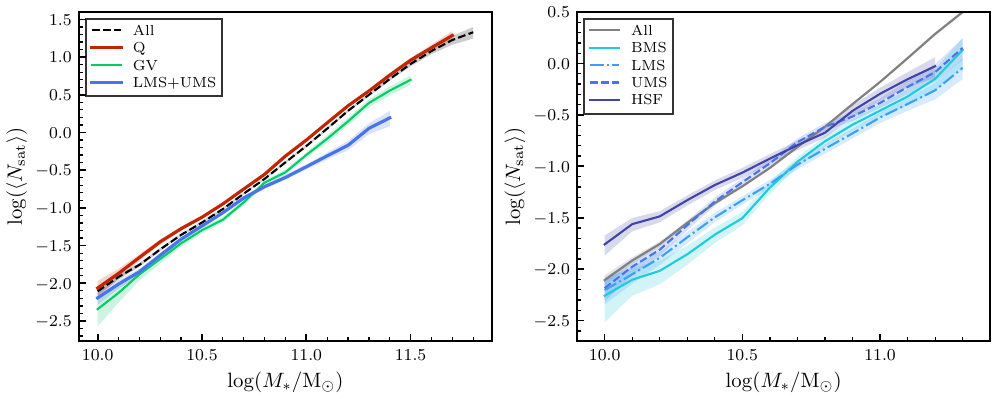}
    \caption{Average number of satellites per central as a function of central galaxy stellar mass. Only mass bins that contain at least 50 centrals are plotted. \textit{Left panel}: The sample is broken into sub-samples of all, LMS+UMS, GV, and quiescent centrals. We see that quiescent centrals on average have more satellites than SFMS centrals, which agrees with our conclusion from the CCFs in Fig.~\ref{fig:CCFs}. At lower masses ($\log(M_*/\Msun)\lesssim10.5$), the average difference is $\sim$0.15~dex, while at higher masses, it grows up to $\sim$0.5~dex. \textit{Right panel}: The SFMS is broken into four sub-samples as a function of $\Delta$MS. Note that the axes have changed to better show the results. For $10.3 \lesssim \log(M_*/\Msun) \lesssim 10.6$, a clear monotonic trend is observed with SFR activity, where higher $\Delta$MS sub-samples have higher $\langle N_\text{sat}\rangle$. For lower masses, the trend is less evident for UMS, LMS, and BMS centrals though HSF centrals clearly have the most satellites, exceeding the average of the entire sample (gray solid line, equivalent to the black dashed line in the left panel). At higher masses, the trend is less clear for all four sub-samples.}
    \label{fig:Nsat}
\end{figure*}

Since the ACFs of central galaxies show little trend with $\Delta$MS, we further investigate the clustering properties of centrals by considering cross-correlations of centrals with satellites. However, we want to do this in a way that will not be influenced by the clustering trends of the satellites. If, for example, we calculate CCFs within each cell of the Fig.~\ref{fig:ssfr-Mstar_plane} grid, we know the clustering amplitudes will be highest in the quiescent cells -- similar to Fig.~\ref{fig:all_ACFs} -- because satellites are highly clustered and tend to be quiescent. To avoid this bias, we instead cross-correlate centrals in a grid cell with \emph{all} satellites in the same mass bin, regardless of satellite sSFR. By doing this, the CCFs will be independent of satellite sSFR and, consequently, unbiased by the clustering trends shown in Figs~\ref{fig:all_ACFs} and \ref{fig:all_ACF_components}.

The central--satellite CCFs are shown in Fig.~\ref{fig:CCFs} using all six $\Delta$MS bins from Fig.~\ref{fig:ssfr-Mstar_plane}. The upper four rows (blue) show results for the SFMS broken into four sections (HSF, UMS, LMS, and BMS), followed by green valley (green) and quiescent (red) centrals. The black dashed lines show the CCFs of centrals in the center of the SFMS (LMS+UMS) within that mass bin with all satellites in the same mass bin. Across the SFMS, we see similar behavior in the CCFs. However, looking at green valley and especially quiescent centrals, there is a general trend of increasing clustering amplitude at smaller separations, i.e., in the one-halo term. For quiescent centrals, the CCF clustering amplitude is greater than that of SFMS centrals by up to a factor of $\sim$3.

Interpreting this increase in clustering is challenging as there are multiple factors at play in the CCFs. First is the projection that is used when calculating the CCFs. Since they are calculated in a projected environment, pair counts of centrals with satellites of another group along the line of sight can lead to higher clustering amplitudes on small scales. If quiescent centrals tend to live in higher density environments, it may be more likely to find cases of quiescent centrals along the line of sight of a larger group than star-forming centrals. Another possibility is pair counts of centrals at the outskirts of larger haloes. Since we calculate the CCFs in mass bins, we are cross-correlating centrals with satellites of similar masses. It is, therefore, likely that satellites of a similar mass to a given central will belong to a different, more massive central. There may then be pair counts from backsplash galaxies \citep[see, e.g.,][]{Borrow+2023}. These galaxies are ones that once resided in a larger cluster but have moved outward on their orbits. In the process of this migration, their gas becomes stripped, resulting in these galaxies tending to be quiescent. A final contributor to the small-scale CCFs is centrals that indeed have satellites with masses that are similar enough to be in the same mass bin. While these cases may be rare, it is possible that they could contribute in a significant way to the small-scale CCFs depending on the frequency of occurrences of the other previously mentioned factors. If this is the case, it would suggest that quiescent centrals tend to have more satellites than star-forming centrals.

To check this conclusion explicitly, we calculate the mean number of satellites per group above a stellar mass of $\log(M_*/\Msun)=10$ for all, SFMS, green valley, and quiescent centrals. For this, we create a new volume-limited sample that spans a mass range of $10 < \log(M_*/\Msun) < 12$ using the same methodology described in Section~\ref{section:data_selection}. The left panel of Fig.~\ref{fig:Nsat} shows the results, plotting only the mass bins that contain at least 50 centrals. As suggested by the CCFs, we see a clear separation in the number of satellites for SFMS and quiescent centrals, with green valley centrals lying generally in the middle at higher central masses. The separation is $\sim$0.15~dex at lower masses, growing up to $\sim$0.5~dex at higher masses. Note that in the halo occupation distribution (HOD) framework, $\langle N_\text{sat}\rangle$ will be proportional to halo mass as $\langle N_\text{sat}\rangle \propto M_\text{vir}^\alpha$ with $\alpha\sim1$ \citep[e.g.,][]{Moster+2010, Zehavi+2011}. Under the assumptions of HOD, where halo mass determines all galaxy properties, the segregation we see in $\langle N_\text{sat}\rangle$ with $\Delta$MS reflects the existence of the SHMR segregation, where quiescent galaxies live in higher mass haloes than star-forming galaxies with the same stellar mass.

The right panel of Fig.~\ref{fig:Nsat} further investigates $\langle N_\text{sat}\rangle$ by dividing the SFMS into the HSF, UMS, LMS, and BMS sub-samples. In the stellar mass range of $10.3 \lesssim \log(M_*/M_{\odot}) \lesssim 10.6$, a clear monotonic trend is observed with SFR activity, where higher $\Delta$MS sub-samples have higher $\langle N_\text{sat}\rangle$. However, for masses below this range, the trend is less evident for UMS, LMS, and BMS. Similarly, for masses above this range, the trend is less distinct for all four sub-samples. For central galaxies with masses $\log(M_*/M_{\odot})\lesssim 10.6$, HSF centrals host more satellites than UMS, LMS, and BMS centrals. Moreover, on average, HSF centrals host $\sim$1.6 times more satellite galaxies compared to the entire sample (gray solid line), while UMS, LMS, and BMS centrals generally host fewer satellites than the overall sample. This trend for HSF centrals aligns with our findings in Fig.~\ref{fig:CCFs}, where the one-halo term of HSF centrals in the lowest mass bin is marginally above that of all SFMS galaxies. The excess of satellite galaxies for HSF centrals may contribute to their elevated SFR values due to close interactions with their satellites (see, e.g., \citealt{Lin+2008, Yesuf2021, Bottrell2024}).

\section{Implications for the galaxy--halo connection} \label{section:galaxy_halo_connection}
With the signal we have seen in the clustering properties of central galaxies -- namely, that more quiescent centrals tend to have more satellites than star-forming centrals (shown in Figs.~\ref{fig:CCFs} and \ref{fig:Nsat}) -- we can now use this to test models of the galaxy--halo connection. To do this, we use the Bolshoi--Planck (BP) $\Lambda$CDM $N$-body simulation \citep{Klypin+2016, Rodriguez-Puebla+2016b}. The BP simulation box is 250~$h^{-1}$Mpc on each side, containing $2048^3$ particles with a mass resolution of $1.55\times10^8~h^{-1}\Msun$ and force resolution of $1.0~h^{-1}$kpc. The cosmological parameters of BP are $\Omega_{\Lambda}=0.693$, $\Omega_\text{m}=0.307$, $\Omega_\text{b}=0.048$, $h=0.678$, $n_s=0.96$, and $\sigma_8=0.823$. Dark matter haloes and sub-haloes in the simulation were identified using \textsc{rockstar} \citep{Behroozi+2013}. Since our correlation functions thus far have been calculated within five different sub-volumes, we use five different sub-volumes created from BP snapshots for these tests. The goal is to have each snapshot cover the same stellar mass and redshift ranges as our five SDSS sub-volumes. To define centrals and satellites in these mocks, we use the designations provided by \textsc{rockstar} in the snapshots. This is the simplest way to make these designations, however it is not necessarily a direct comparison to observations. As shown by \citet{Campbell+2015}, color-dependent group statistics can be affected in non-trivial ways by group finder errors. This mainly occurs when the analysis is performed as a function of halo mass, due to errors in halo mass assignments, and to a lesser degree when using the luminosity/mass of the central galaxy. Since we are not employing halo masses from any of the group finders, we do not consider the above to be a serious source of uncertainty (see Section~\ref{section:group_finder_errors} for details). Nonetheless, for a more direct comparison between the mocks and observations, one could run different group finder methods on the mocks to define centrals and satellites. We leave this task to a future paper (Enriquez-Vargas et al. in preparation).

\begin{table*}
    \centering
    \caption{Description of the three sSFR models of the galaxy--halo connection. The correlation direction tells whether a given halo property correlates with sSFR positively (higher halo property corresponds to higher sSFR) or negatively (higher halo property corresponds to lower sSFR).}
    \begin{tabularx}{\linewidth}{c c X}
        \hline
        \rule{0pt}{8pt}Halo Property & Correlation with sSFR & \multicolumn{1}{c}{Motivation} \\
        \hline
        \rule{0pt}{10pt}Halo accretion rate ($\dot{M}_\text{h}$) & Positive & Halo accretion rate controls the influx of gas into a galaxy's interstellar medium. We might expect galaxies in haloes with higher accretion rates may have higher sSFR as they have more gas available to form stars. \\
        \rule{0pt}{12pt}Halo concentration ($C_\text{vir}$) & Negative & Halo concentration is linked to the timing of gas infall into haloes where more concentrated haloes experience earlier infall. We might expect galaxies in highly concentrated haloes have low sSFR as they have run out of gas and ceased star formation. \\
        \rule{0pt}{12pt}Peak circular velocity ($V_\text{peak}$) & Negative & $V_\text{peak}$ is strongly correlated with halo mass. If quiescent galaxies reside in more massive haloes, we can assign lower sSFRs to haloes with higher $V_\text{peak}$. \rule[-6pt]{0pt}{0pt} \\
        \hline 
    \end{tabularx}
    \label{tab:galhalo_models}
\end{table*}

\subsection{Mock catalogues: redshift space and stellar masses}
To create our five samples using BP, we first project snapshots into redshift space. We start by defining an origin that is centered on one side of the simulation box. Assuming the snapshot corresponds to a redshift $z_\text{BP}$, we shift all haloes away from the origin a distance $d_\text{c}(z_\text{BP})$, where $d_\text{c}(z_\text{BP})$ is the comoving distance to $z_\text{BP}$. This sets the redshift of the near side of the box equal to $z_\text{BP}$. Then, we infer cosmological redshifts for all haloes moving across to the far side of the box based on their comoving distances from the origin. Additionally, we add redshift-space distortions according to the equation
\begin{equation}
    z_\text{obs} = z_\text{cos} + \frac{v_\text{los}}{c}\parenth{1+z_\text{cos}}
\end{equation}
where $z_\text{obs}$ is the observed redshift with redshift-space distortions, $z_\text{cos}$ is the cosmological redshift inferred from a halo's comoving distance from the origin, $v_\text{los}$ is the line-of-sight peculiar velocity of a halo relative to the origin, and $c$ is the speed of light. Once every halo has an observed redshift assigned to it, we then cut the snapshots so each one has the same redshift range as the corresponding sub-volume in our SDSS samples. In some cases, the snapshot is not large enough to encompass the entire redshift range of a sub-volume. For these instances, we choose BP snapshots that give us a redshift range that is as close as possible to our SDSS sub-volume redshift ranges.\footnote{For the BP snapshots we use, all redshift boundaries are within 0.01 of the corresponding SDSS sub-volume redshift boundaries.}

Once all the snapshots and redshift ranges have been determined, we assign stellar masses to all of the haloes. To do this, we follow \citet[][see also \citealp{Reddick+2013, Calette+2021}]{Dragomir+2018} and assume that the halo property that best correlates with stellar mass and reproduces the observed ACF is
\begin{equation}
    V_\text{DM} = 
    \begin{cases}
        V_\text{max} & \text{for distinct haloes}\\
        V_\text{peak} & \text{for sub-haloes}
    \end{cases}
    \label{eq:vhmr}
\end{equation}
where $V_\text{max}$ is the maximum circular velocity of dark matter in a distinct halo at the observed time and $V_\text{peak}$ is the maximum circular velocity throughout the entire history of a sub-halo. Moreover, we follow \citet*[][see also \citealp{Rodriguez-Puebla+2013, Guo+2016, Chandrachani_Devi+2019}]{Rodriguez-Puebla+2012} to separately derive $V_\text{DM}$--$M_*$ relations for centrals and satellites. As shown in \citet{Rodriguez-Puebla+2012}, using separate relations for centrals and satellites results in ACFs that are more consistent with observations than using the same relation for both.

We follow the same procedure described in Section~2.2 of \citet{Calette+2021} to derive the $V_\text{DM}$--$M_*$ relations based on the numerical deconvolution algorithm discussed in Appendix~D of \citet{Rodriguez-Puebla+2020b}. One difference with respect to \citet{Calette+2021} is that we assume that their equation~(3) can be written separately for centrals and satellites as
\begin{equation}
    \phi_{i}(M_*, z) = \iint \frac{\mathcal{H}_i(M_*|V_\text{DM}) \phi_{\text{DM},i}(V_\text{DM})} {\mathcal{V}_\text{c}(z_f) - \mathcal{V}_\text{c}(z_i)} \mathrm{d}\log(V_\text{DM}) \mathrm{d}\mathcal{V}_\text{c}(z),
    \label{eq:galaxy_halo_connection}
\end{equation}
where the subscript $i$ refers to either centrals/distinct haloes or satellites/sub-haloes, $\mathcal{V}_\text{c}$ is the comoving volume, and $z_i = 0.001$ and $z_f = 0.2$ correspond to the redshift ranges over which the GSMF that we use has been constrained. The function $\phi_{\text{DM},i}(V_\text{DM})$ is the halo/sub-halo velocity function and $\mathcal{H}_i(M_*|V_\text{DM})$ refers to the log-normal conditional probability distribution function that a halo/sub-halo with velocity $\log(V_\text{DM}) \pm \frac12\mathrm{d}\log(V_\text{DM})$ hosts a galaxy with a mass $\log(M_*) \pm \frac12\mathrm{d}\log(M_*)$. We assume that the dispersion around the $V_\text{DM}$--$M_*$ relation is $\sigma=0.15$~dex and the same for both central and satellite galaxies. Note that equation~(\ref{eq:galaxy_halo_connection}) returns the volume-weighted $V_\text{DM}$--$M_*$ relation of centrals and satellites.

To derive the $V_\text{DM}$--$M_*$ relationship, we use the GSMF of all galaxies, $\phi(M_*)$, from \citet{Dragomir+2018}. We calculate the GSMF of central and satellite galaxies by computing the fraction of satellites, $f_{\text{sat}}$, as a function of stellar mass using the volume-limited samples described in Section~\ref{section:data}. The GSMF of central and satellite galaxies will then, by definition, be given by $\phi_{i}(M_*) = f_{i} \phi(M_*)$ with $f_{\text{cen}} = 1- f_{\text{sat}}$. Using $\phi_{i}(M_*)$ on the left side of equation~(\ref{eq:galaxy_halo_connection}) allows us to deconvolve the equation to isolate $\mathcal{H}_i(M_*|V_\text{DM})$ and calculate stellar masses for all the dark matter haloes. Once the haloes have stellar masses assigned to them, we select our final BP samples by choosing the corresponding stellar mass and redshift ranges to match our five SDSS sub-volumes.

\subsection{Assigning sSFRs to dark matter haloes}
We now test different models of sSFR for central galaxies. For this step, we consider only centrals/distinct haloes as we expect satellites/sub-haloes to have different formation and evolutionary histories from the host haloes in which they reside. Note that we model sSFR empirically using our snapshots and SDSS observations at $z\leq0.2$; we do not attempt to model the histories of the haloes. For a given model, we choose a halo property, $H$ (e.g., $C_\text{vir}$), with which sSFR will correlate as either a monotonic increasing or decreasing relationship.\footnote{These are the most straightforward trends that have a physical basis for sSFR to be correlated with halo property $H$.} Then, in bins of stellar mass, we rank order all haloes by $H(M_*)$. Assuming there are $N$ haloes in a given mass bin, we sample $N$ sSFRs from the corresponding mass bin of SDSS observations. These sampled sSFRs are also rank ordered and then matched to the haloes based on rank (i.e., if $H$ and sSFR are positively correlated, the highest sampled sSFR will be assigned to the halo with the highest value of $H(M_*)$).

Using this procedure, we test three different models of sSFR: halo accretion rate\footnote{Since we are modeling \emph{specific} SFR, it is also sensible to correlate this with specific halo accretion rate ($\dot{M}_\text{h}/M_\text{h}$). We tested this model as well and found similar results to the $\dot{M}_\text{h}$ model.} (averaged over a dynamical time), $\dot{M}_\text{h}$, halo concentration, $C_\text{vir}$, and peak circular velocity, $V_\text{peak}$, which are summarized in Table~\ref{tab:galhalo_models}. As mentioned in Section~\ref{section:introduction}, halo accretion rate controls the influx of gas into a galaxy's interstellar medium \citep[see, e.g.,][]{Avila-Reese+2000, Bouche+2010, DekelMandelker2014, Rodriguez-Puebla+2016a} and has been shown to reproduce the width of the SFR distribution in star-forming galaxies at various redshifts \citep{Rodriguez-Puebla+2016a} and, to a lesser extent, the clustering of galaxies \citep{Becker+2015}. Previous studies of $C_\text{vir}$ have shown that it is linked to the timing of gas infall into haloes \citep[see, e.g.,][]{Avila-Reese+1998, Wechsler+2002, Gao+2004, Dutton+2010a, Matthee+2019}, and models of galaxy color based on $C_\text{vir}$ have successfully reproduced the observed ACFs of red and blue galaxies \citep{Hearin_Watson_2013}. 

The $\dot{M}_\text{h}$ and $C_\text{vir}$ models will be used as tests of whether halo assembly history is the key factor determining how galaxies cluster in the sSFR--$M_*$ plane. At a fixed halo mass, haloes with low $\dot{M}_\text{h}$ will tend to be the ones in denser environments comprised of already-formed haloes \citep[see, e.g.,][]{Maulbetsch+2007, Lee+2017} and sub-haloes with less dark matter available for accretion. We might expect that the lower accretion rate leads to less gas available to a halo's hosted galaxy and therefore a lower sSFR. In the $\dot{M}_\text{h}$ model, we then expect quiescent galaxies to be more clustered than star-forming galaxies at a fixed stellar mass. Similarly, at a fixed halo mass, haloes with high $C_\text{vir}$ will tend to be the ones that formed at earlier times and are more clustered. We then might expect that the galaxies hosted by such haloes also formed earlier and reached the ends of their life cycles (i.e., became quiescent) earlier. In the $C_\text{vir}$ model, we can then expect that quiescent galaxies will again be more clustered than star-forming galaxies at a fixed stellar mass.

Contrary to these models, the $V_\text{peak}$ model will be a test that assumes only a segregation in the SHMR. $V_\text{peak}$ is strongly correlated with halo mass; therefore, assigning lower sSFR to haloes with higher $V_\text{peak}$ will lead to quiescent galaxies occupying more massive haloes than star-forming galaxies at the same stellar mass. It is important to note that the motivation for this model is empirical according to the results from gravitational weak lensing \citep{Mandelbaum+2006}, galaxy groups and clustering \citep{Tinker+2013, Rodriguez-Puebla+2015}, and galaxy kinematics \citep{More+2011}, as discussed in Section~\ref{section:introduction}. In this way, the difference in clustering for quiescent and star-forming centrals will be driven by halo bias. This could be tested more directly with a model based on halo mass rather than $V_\text{peak}$. However, we tested such a model and found that $V_\text{peak}$ overall provided results more consistent with observations.

\subsection{Clustering predictions}
\begin{figure*}
    \centering
    \includegraphics[width=\linewidth]{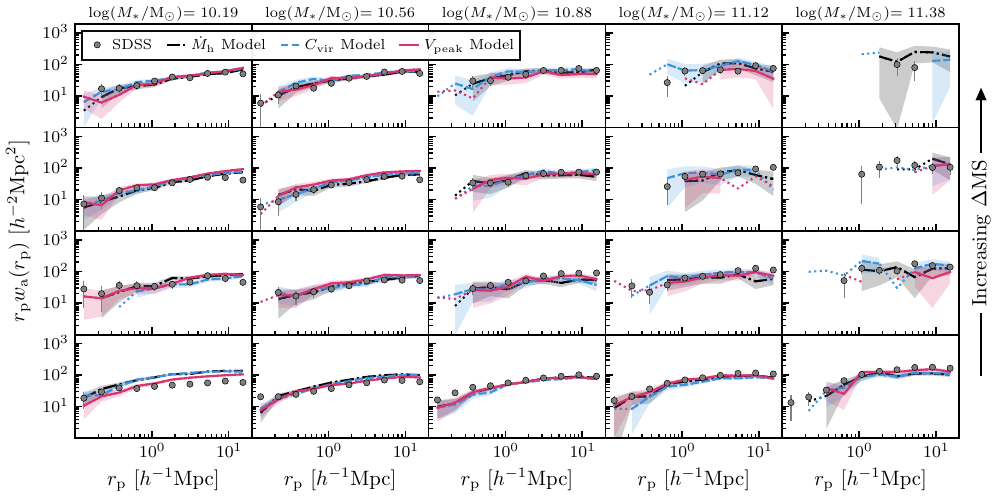}
    \caption{Projected auto-correlation functions (ACFs) of central galaxies as a function of $\Delta$MS and $M_*$. The rows correspond from top to bottom to UMS+HSF, BMS+LMS, GV, and quiescent centrals. The gray circles show results for SDSS observations and the lines show results of three different models of sSFR based on different halo properties using the Bolshoi--Planck simulation. The dash-dot black lines show the $\dot{M}_\text{h}$ model, the dashed blue lines show the $C_\text{vir}$ model, and the solid magenta lines show the $V_\text{peak}$ model. Dotted lines show results (without error bars) where the error reaches up to $2w_\text{a}$. We see that all models are consistent with SDSS and produce ACFs that are very similar.}
    \label{fig:galhalo_ACFs}
\end{figure*}
\begin{figure*}
    \centering
    \includegraphics[width=\linewidth]{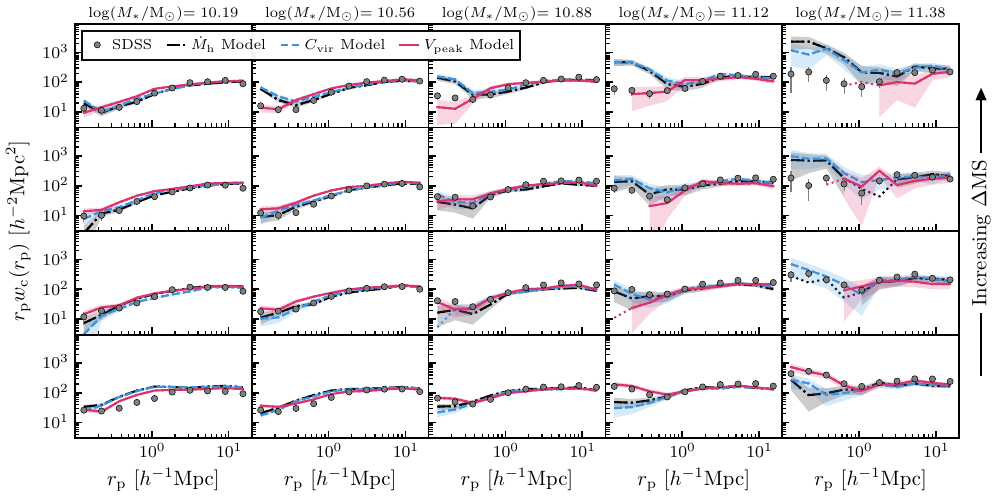}
    \caption{Projected cross-correlation functions (CCFs) of centrals in a given grid cell with all satellites in the same mass bin. The rows correspond from top to bottom to UMS+HSF, BMS+LMS, GV, and quiescent centrals. The gray circles show results for SDSS observations and the lines show results of three different models of sSFR based on different halo properties using the Bolshoi--Planck simulation. The dash-dot black lines show the $\dot{M}_\text{h}$ model, the dashed blue lines show the $C_\text{vir}$ model, and the solid magenta lines show the $V_\text{peak}$ model. Dotted lines show results (without error bars) where the error reaches up to $2w_\text{c}$. We see that the $\dot{M}_\text{h}$ and $C_\text{vir}$ models for sSFR fail to reproduce the CCF trend in $\Delta$MS that we find in observed data. At higher masses, clustering amplitude is highest for star-forming centrals, disagreeing strongly with SDSS. However, the $V_\text{peak}$ model is consistent with SDSS at all masses.}
    \label{fig:galhalo_CCFs}
\end{figure*}
\begin{figure}
    \centering
    \includegraphics[width=0.8\linewidth]{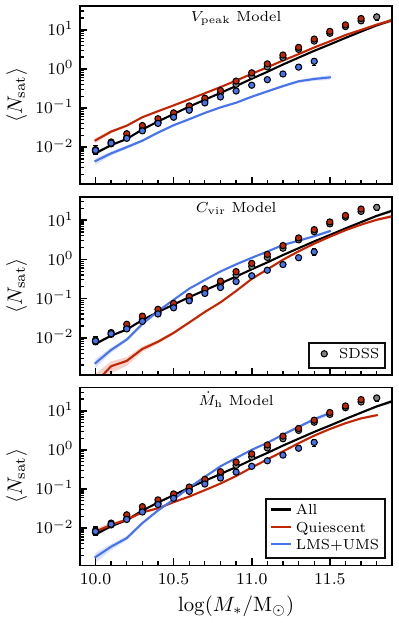}
    \caption{Average number of satellites per central as a function of central galaxy stellar mass. Only mass bins that contain at least 50 centrals are plotted. The sample is broken into sub-samples of all, LMS+UMS, and quiescent centrals. The circles show results for SDSS observations and the lines show results of three different models of sSFR based on different halo properties using the Bolshoi--Planck simulation. The $C_\text{vir}$ model predicts that star-forming centrals have a greater $\langle N_\text{sat}\rangle$ than quiescent centrals, contrary to the observations. This also occurs in the $\dot{M}_\text{h}$ model above  $\log(M_*/\Msun) \approx 10.4$ where the $\langle N_\text{sat}\rangle$ relation inverts. The $V_\text{peak}$ model, on the other hand, agrees with the observations that quiescent centrals have more satellites than star-forming centrals at all masses.}
    \label{fig:galhalo_Nsat}
\end{figure}
For each of the models, we calculate ACFs and CCFs using the same methodology described in Sections~\ref{section:methods} and \ref{section:results}. Note that because we cross-correlate centrals with all satellites in the same mass bin regardless of satellite sSFR, these results do not depend on a model of sSFR for satellites. We show the results of the ACFs and CCFs for the $\dot{M}_\text{h}$, $C_\text{vir}$, and $V_\text{peak}$ models in Figs~\ref{fig:galhalo_ACFs} and \ref{fig:galhalo_CCFs}, respectively. The rows correspond to HSF+UMS, LMS+BMS, GV, and quiescent centrals from top to bottom. We plot the ACFs and CCFs of SDSS observations as circles and the models as lines.

We first consider the ACFs of the three models, shown in Fig.~\ref{fig:galhalo_ACFs}. The predicted ACFs agree fairly well with the SDSS observations, showing little to no trend with $\Delta$MS at a fixed stellar mass. This result is initially counterintuitive for all models. In the case of the $\dot{M}_\text{h}$ and $C_\text{vir}$ models, it may be expected that the assembly bias effects built into them would result in quiescent galaxies clustering more strongly than star-forming galaxies of the same stellar mass. For the $V_\text{peak}$ model, we effectively assign lower sSFRs to higher mass haloes at the same stellar mass, leading to the expectation that halo bias would cause quiescent galaxies to cluster more strongly than star-forming galaxies at a given stellar mass. However, neither of these effects is observed. To understand this result, one must consider the behavior of halo bias as a function of halo mass. The magnitude of halo bias is a function of halo mass and ramps up significantly at higher masses around $\log(M_\text{h}/\Msun)\gtrsim13.5$ (see, e.g., \citealp{Wechsler_Tinker2018}). In the $V_\text{peak}$ model, the highest halo masses for star-forming and quiescent galaxies are, on average, around $\log(M_\text{h}/\Msun)\sim13$ and $\log(M_\text{h}/\Msun)\sim13.5$, respectively (see Section~\ref{section:shmr}). At these halo masses, the difference in halo bias is not significant enough to produce a signal in the auto-correlations. For the $\dot{M}_\text{h}$ and $C_\text{vir}$ models, the highest halo masses for star-forming and quiescent galaxies are around $\log(M_\text{h}/\Msun)\sim13.9$ and $\log(M_\text{h}/\Msun)\sim13.4$, respectively. Note that these models predict that quiescent galaxies reside in \emph{lower} mass haloes than star-forming galaxies at fixed stellar mass. Because these halo masses are higher, the difference in halo bias is stronger. It may then be the case that halo bias increases the clustering of star-forming galaxies while assembly bias increases the clustering of quiescent galaxies, resulting in no net difference in the ACFs as a function of $\Delta$MS. It is also worth noting that assembly bias is an effect at fixed \emph{halo} mass, and it is not straightforward how it translates when working at fixed \emph{stellar} mass as we are here.

\begin{figure*}
    \centering
    \includegraphics[width=0.9\linewidth]{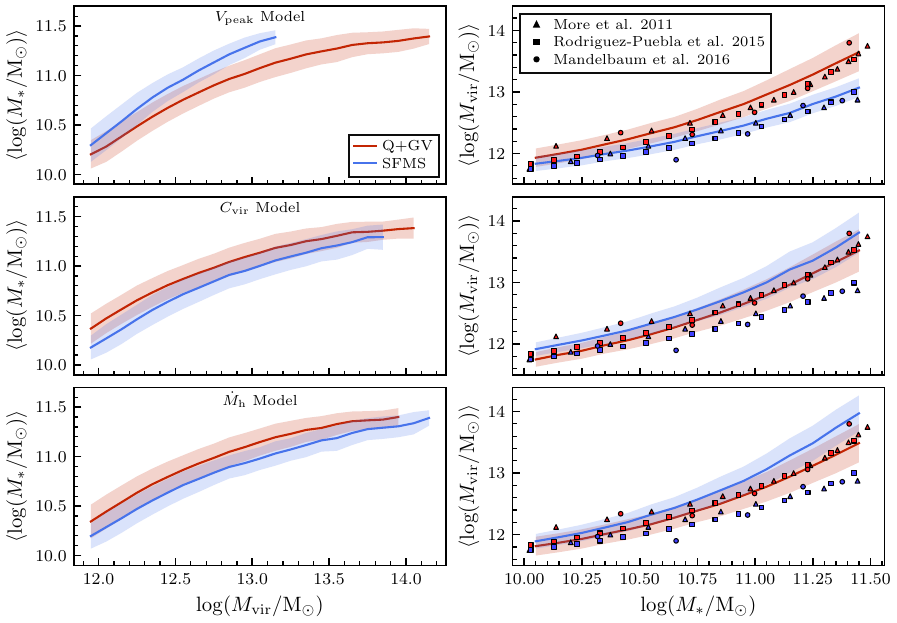}
    \caption{The stellar-to-halo mass relation (SHMR), $\langle M_*(M_\text{vir})\rangle$ (left panels), and the inverted SHMR $\langle M_\text{vir}(M_*)\rangle$ (right panels), for quiescent and green valley (Q+GV) and SFMS centrals using the $V_\text{peak}$, $C_\text{vir}$, and $\dot{M}_\text{h}$ models of sSFR. Only mass bins with at least 50 galaxies are plotted for the different models. Also shown in the right panels are determinations using satellite kinematics \citep{More+2011}, galaxy groups combined with galaxy clustering \citep{Rodriguez-Puebla+2015}, and weak lensing \citep{Mandelbaum+2016}. We see that while all models produce inverted SHMRs that agree with observations for Q+GV galaxies, only the $V_\text{peak}$ model reproduces the observed relation for SFMS galaxies. We also note that our model results are not affected by the inversion problem.}
    \label{fig:shmr}
\end{figure*}

Looking now at the CCFs of the $\dot{M}_\text{h}$ and $C_\text{vir}$ models in Fig.~\ref{fig:galhalo_CCFs}, we see that these models produce very similar results. They agree well with the observations at the lowest masses, aside from the quiescent centrals where they predict higher clustering in the one-halo term. In the second-lowest mass bin, there is an upturn in clustering amplitude in the one-halo term for the higher star-forming centrals. This upturn grows with stellar mass and disagrees strongly with what we observed in the SDSS. As mass increases, the $\dot{M}_\text{h}$ and $C_\text{vir}$ models produce a trend with $\Delta$MS \emph{opposite} of what we see in observations; clustering is highest for the star-forming centrals and lowest for quiescent centrals. 

In contrast to the $\dot{M}_\text{h}$ and $C_\text{vir}$ models, the $V_\text{peak}$ model CCFs produce reasonable agreement with the SDSS observations at all masses but especially in the three higher mass bins. At lower masses, the SFMS clustering has a tendency to be over-predicted in the one-halo term. However, the main success of this model is that it generally shows increased clustering amplitude for quiescent galaxies, which agrees with the trend in the observations, unlike what the $\dot{M}_\text{h}$ and $C_\text{vir}$ models predicts.

An important thing to note from these correlation function results is the fact that the CCFs showed differing results among the models, whereas the ACFs did not. In almost every region of our sSFR--$M_*$ plane, all three models predict ACFs that are nearly the same while the $V_\text{peak}$ model was able to clearly distinguish itself via CCFs. What this tells us is that when considering only central galaxies, ACFs alone may not be sufficient to discriminate different models of galaxy sSFR. We do add, however, that since the presence of satellites has a strong impact on clustering trends with sSFR, it is possible that ACFs of \emph{all} galaxies (i.e., centrals and satellites together) could differentiate models of galaxy sSFR, though we are unable to test this given that we do not attempt to model sSFRs for sub-haloes.

Finally, we consider the predictions of $\langle N_\text{sat}\rangle$ by the three models, shown in Fig.~\ref{fig:galhalo_Nsat}. In the SDSS observations, there is a consistent trend at all masses of quiescent galaxies having more satellites than star-forming galaxies, and this trend is weakest at lower masses and strongest at higher masses. In these regards, we again see clear failures in the $\dot{M}_\text{h}$ and $C_\text{vir}$ models. While the $\dot{M}_\text{h}$ model does predict lower $\langle N_\text{sat}\rangle$ for star-forming galaxies at lower masses, there is a reversal of this trend above $\log(M_*/\Msun)\approx10.4$, contrary to what the observations show. Additionally, the $\langle N_\text{sat}\rangle$ values for quiescent and star-forming galaxies are diverging at lower masses rather than converging as the observations do. The $C_\text{vir}$ model incorrectly predicts that star-forming galaxies have greater $\langle N_\text{sat}\rangle$ at all masses, opposite of what the observations show, and has converging values of $\langle N_\text{sat}\rangle$ for quiescent and star-forming galaxies at both high and low masses. On the other hand, the $V_\text{peak}$ model agrees with the observations in its prediction that quiescent galaxies have greater $\langle N_\text{sat}\rangle$ than star-forming galaxies at all masses, with the difference being the greatest at higher masses. The failures of the $\dot{M}_\text{h}$ and $C_\text{vir}$ models show that conventional assembly bias models of sSFR do not reproduce the observed clustering of central galaxies. The successes of the $V_\text{peak}$ model, however, show that the clustering of central galaxies in the sSFR--$M_*$ plane can be largely explained by a mass-based model, independent of assembly bias.

\begin{figure*}
    \centering
    \includegraphics[width=0.75\linewidth]{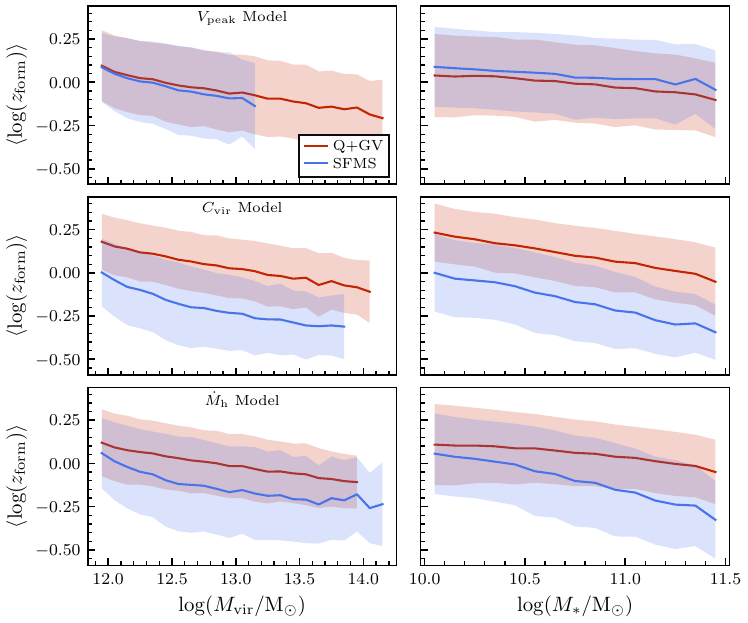}
    \caption{Formation redshift, $z_\text{form}$, as a function of halo mass (left panels) and stellar mass (right panels) for three models of sSFR broken into quiescent and green valley (Q+GV) and SFMS centrals. Formation redshift is defined as the redshift at which a halo attained 50\percent{} of its current mass. The $C_\text{vir}$ and $\dot{M}_\text{h}$ models each create a strong segregation in formation redshift between Q+GV and SFMS galaxies. This is consistent with the expected behavior of the assembly bias effect that is built into these models, where early-forming haloes host quiescent galaxies. In contrast, the $V_\text{peak}$ model has no clear bias in formation redshift with sSFR, which supports our hypothesis that the $V_\text{peak}$ model is strongly associated with halo mass.}
    \label{fig:formation_redshift}
\end{figure*}

\subsection{Stellar-to-halo mass relations} \label{section:shmr}
In this section, we analyze the SHMR predictions of the three models. Fig.~\ref{fig:shmr} shows the resulting SHMRs in the left panels and the inverted SHMRs in the right panels for each of the three models described in the preceding sections. We note that none of the three models is affected by the inversion problem \citep{Cui+2021}. In each panel, the blue and red lines represent the SFMS centrals and a sample consisting of quiescent and green valley centrals (Q+GV), respectively. When using $V_\text{peak}$ (top panels), SFMS galaxies exhibit a larger stellar mass than Q+GV galaxies at a fixed halo mass (left panel). This outcome is a consequence of assigning increasing values of sSFR to decreasing values of $V_\text{peak}$, as illustrated in the right panel of Fig.~\ref{fig:shmr}. Since $V_\text{peak}$ strongly correlates with halo mass, this effectively results in assigning quiescent galaxies to more massive haloes than star-forming galaxies at the same stellar mass (consequently, quiescent galaxies will then have lower stellar masses than star-forming galaxies at the same halo mass). The $C_\text{vir}$ and $\dot{M}_\text{h}$ models (middle and bottom panels) result in opposite trends to the $V_\text{peak}$ model, both in the SHMR and the inverted SHMR. These trends can be understood as follows. In the $C_\text{vir}$ model, we assign increasing values of sSFR to decreasing values of concentration at a fixed stellar mass. Since at $z\sim0$ halo concentration correlates with halo mass as $C_\text{vir}\propto M_\text{vir}^{-0.1}$ \citep{Maccio+2008, Klypin+2016}, we are effectively assigning star-forming galaxies to high-mass haloes. For the $\dot{M}_\text{h}$ model, increasing values of sSFR were assigned to increasing values of halo accretion rate at a given stellar mass. Since $\dot{M}_\text{h}\propto M_\text{vir}^{1.1}$ \citep{Fakhouri+2010, Rodriguez-Puebla+2016b}, we are again effectively assigning star-forming galaxies to high-mass haloes.

Our findings reveal a distinct segregation in the SHMR when considering SFMS and Q+GV centrals across the three models of sSFR. When compared to earlier determinations using satellite kinematics \citep{More+2011}, galaxy groups combined with galaxy clustering \citep{Rodriguez-Puebla+2015}, and weak lensing \citep{Mandelbaum+2016}, we observe that all models agree for Q+GV galaxies, but only the $V_\text{peak}$ model aligns with the trend where quiescent galaxies inhabit more massive haloes compared to their star-forming counterparts.

\subsection{Caveats and interpretations}
In this section, we briefly address some caveats that may affect some of the results of our models. Regarding the $C_\text{vir}$ model, there may be some fraction of haloes that exhibit low values of $C_\text{vir}$ because they have not yet relaxed after mergers. These merging haloes are typically found in higher density environments and are expected to host quiescent galaxies. However, due to their temporarily low $C_\text{vir}$, they could be erroneously assigned high sSFR values, which might contribute to the unexpected trends seen in Figs~\ref{fig:galhalo_CCFs} and \ref{fig:galhalo_Nsat}. As for the $\dot{M}_\text{h}$ model, in some instances, high mass accretion rates may involve major mergers, particularly at high masses where the halo should actually host a quiescent central galaxy. In the model, however, such haloes would be assigned high sSFR. This scenario may also contribute to the unexpected trends in Figs~\ref{fig:galhalo_CCFs} and \ref{fig:galhalo_Nsat}. While these caveats may impact the results quantitatively, we expect the qualitative aspects would remain unchanged.

Turning now to the $V_\text{peak}$ model, one might initially assume that the sSFR segregation imposed in the $V_\text{peak}$--$M_*$ relation could be attributed, in part, to assembly bias related to $V_\text{peak}$. This assumption arises from the close relationship between $V_\text{peak}$ and $C_\text{vir}$ \citep[see, e.g.,][]{Klypin+2016}, where higher values of $C_\text{vir}$ will generally correspond to higher values of $V_\text{peak}$ for a given halo mass. However, as shown in Fig.~\ref{fig:shmr}, this may not necessarily hold true, as the segregation of the SHMR differs between the two models.

To investigate this further, we consider the formation redshift, $z_\text{form}$, for haloes hosting SFMS and Q+GV galaxies in our models, where $z_\text{form}$ is defined as the redshift at which a dark matter halo reached 50\percent{} of its current mass. Fig.~\ref{fig:formation_redshift} presents $z_\text{form}$ as a function of halo mass (left panels) and stellar mass (right panels). The top row shows the results of the $V_\text{peak}$ model, where haloes hosting SFMS and Q+GV galaxies exhibit minimal difference in $z_\text{form}$ as a function of halo or stellar mass. In contrast, both the $C_\text{vir}$ and $\dot{M}_\text{h}$ models demonstrate a strong correlation of sSFR with $z_\text{form}$. This is consistent with the expected behavior of halo assembly bias, where early-forming haloes tend to host quiescent galaxies. These results support our initial hypothesis that the $V_\text{peak}$ model is significantly associated with halo mass.

We conclude that the empirical evidence obtained in this section \emph{strongly} suggests that the $V_\text{peak}$ model, in which the enhanced clustering of quiescent galaxies is attributed to the fact that they occupy more massive haloes, offers a realistic model of how normal to massive central galaxies (i.e., $\log(M_*/\Msun)>10$) inhabit dark matter haloes.

\section{Discussion} \label{section:discussion}
In this paper, we have investigated how the clustering properties of galaxies vary across the sSFR--$M_*$ plane. To do this, we defined five volume-limited sub-samples that are complete in stellar mass within their respective mass bins, as outlined in Section~\ref{section:data_selection}. Additionally, we confirmed that our sub-samples reproduce the SDSS GSMF \citep{Dragomir+2018}, as shown in Fig.~\ref{fig:gsmf}.

For each of these five sub-samples, we analyzed the ACFs of all galaxies together (Fig.~\ref{fig:all_ACFs}) as well as individually for centrals (Fig.~\ref{fig:centrals_ACFs}) and satellites, all of which are summarized in Fig.~\ref{fig:all_ACF_components}. Our analysis revealed that the ACFs of all galaxies strongly depend on sSFR at a given stellar mass, while the ACFs of central galaxies are largely independent of sSFR. We concluded that the presence of satellite galaxies and the fraction of satellites as a function of sSFR play a crucial role in these differences.

Next, we explored the cross-correlation of central galaxies with satellites across the sSFR--$M_*$ plane (Fig.~\ref{fig:CCFs}). We demonstrated that there is a consistent trend in all mass bins of increased clustering amplitude in the one-halo term for lower $\Delta$MS at a fixed stellar mass. The clustering amplitude of quiescent centrals was as much as $\sim$3 times greater than that of SFMS centrals. This observation aligns with the expectation that quiescent central galaxies tend to have more satellites than those on the SFMS. We supported this observation by showing a difference in the average number of satellites for SFMS and quiescent central galaxies in Fig.~\ref{fig:Nsat}. We found that the average number of satellites for quiescent centrals is about 2--3 times larger than that of SFMS centrals at higher masses.

In the following sections, we discuss the robustness of our results and comparisons with the literature.

\subsection{Robustness of results} \label{section:robustness_of_results}
\begin{table}
    \centering
    \caption{Per cent agreement of central and satellite designations across all three group catalogues in our five sub-volumes.}
    \begin{tabular}{c c}
        \hline
        \rule{0pt}{8pt}$M_*$ Limits $\bracket{\log(M_*/\Msun)}$ & Agreement \\
        \hline
        \rule{0pt}{8pt}\phantom{00}10.0 -- 10.375 & 68.3\% \\
        10.375 -- 10.75\phantom{0} & 72.8\% \\
        \phantom{0}10.75 -- 11.0\phantom{00} & 78.3\% \\
        \phantom{00}11.0 -- 11.25\phantom{0} & 84.4\% \\
        \phantom{0}11.25 -- 11.5\phantom{00} & 90.1\% \\
        \hline
    \end{tabular}
    \label{tab:group_catalog_match}
\end{table}

\begin{figure}
    \centering
    \includegraphics[width=\linewidth]{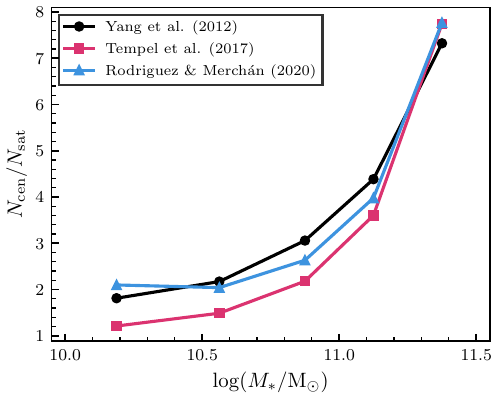}
    \caption{Ratios of central to satellite galaxies in each mass bin for each of the three group catalogues. \citetalias{Tempel+2017} generally has the lowest ratio, i.e., the most galaxies designated as satellites. This is useful for testing central-only samples as one would expect \citetalias{Tempel+2017} to have a more conservative estimate of centrals, leaving a sample with less contamination from satellites misidentified as centrals.}
    \label{fig:centrals_per_satellite}
\end{figure}

\begin{figure*}
    \centering
    \includegraphics[width=\linewidth]{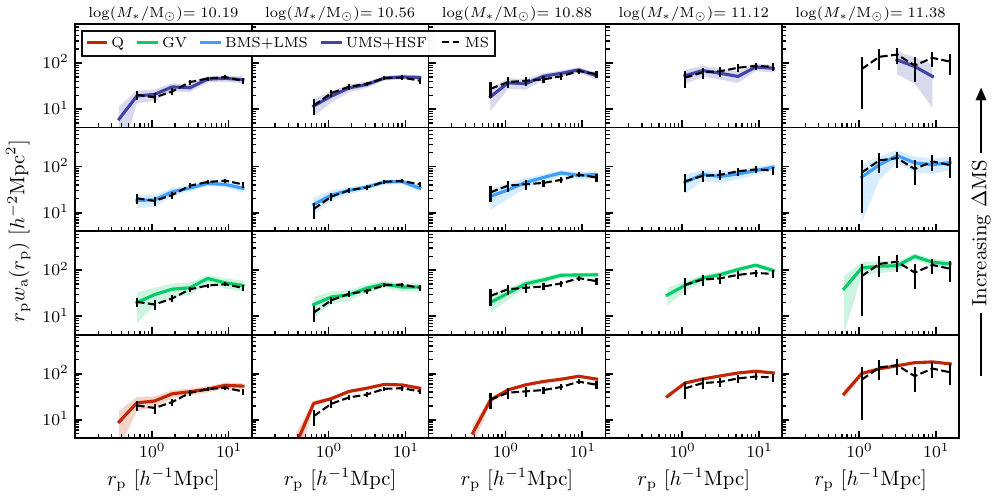}
    \caption{Projected auto-correlation functions (ACFs) of central galaxies identified by \citetalias{Tempel+2017} as a function of $\Delta$MS and $M_*$. The rows correspond from top to bottom to UMS+HSF galaxies (dark blue), BMS+LMS galaxies (light blue), GV galaxies (green), and quiescent galaxies (red). The colored lines show the results within a grid cell, and the black dashed lines show the ACFs of central galaxies in the center of the SFMS (LMS+UMS). Similar to \citetalias{Yang+2012}, there is no strong trend in the ACFs as a function of $\Delta$MS, though there is a slight bias to higher clustering for the quiescent centrals. Unlike \citetalias{Yang+2012}, there is no signal in the quiescent centrals above the SFMS ACFs at small separations. See Section~\ref{section:group_catalog_comparison} for details on this.}
    \label{fig:tempel_centrals_ACFs}
\end{figure*}

In this section, we will discuss comparisons of the group catalogues used and the results obtained from them. Since the results of this paper depend on the identification of central and satellite galaxies, it is important to consider the same analysis using other group catalogues. While it is not possible to directly check the accuracy of any given group catalogue, checking for systematic trends in our results among different catalogues will test the robustness of our findings.

\subsubsection{Group finder systematic errors} \label{section:group_finder_errors}
Before comparing different group finders, it is important to acknowledge the extensive literature discussing the errors in group finding algorithms \citep[see, e.g.,][and references therein]{Duarte+2014, Campbell+2015, Duarte+2015}. As discussed in \citet{Campbell+2015}, systematic errors in group finders arise due to inaccuracies in determining group membership and in designating central and satellite galaxies. One of the most significant challenges is the central/satellite designation, which affects the purity and completeness of these categories as a function of group mass.

In the \citetalias{Yang+2012} group catalogue, a central galaxy is defined as the brightest galaxy in a group. Here, we have redefined it as the most \emph{massive} galaxy within a group. To understand the impact of this assumption, consider the case of a perfect group finding algorithm, where group members are correctly identified but the central/satellite designations may be incorrect due to the most massive group member not being the true central. According to \citet{Campbell+2015}, these errors mostly affect massive groups/clusters ($M_* > 5\times10^{13}\Msun$). It is well known that in dense environments such as these massive groups/clusters, the fraction of quenched and early-type galaxies is higher compared to galaxies in more isolated environments \citep{Dressler1980, Balogh+2004, Peng+2010, Einasto+2024}. This trend is also observed as a function of halo mass \citep{Weinmann+2006}. Additionally, the fraction of quenched and early-type galaxies increases towards the center of the cluster \citep{Weinmann+2006} and is higher when the central galaxy is quiescent/early-type, known as galactic conformity \citep{Weinmann+2006, Kauffmann+2013, Otter+2020, Ayromlou+2023}. Moreover, the luminosity/mass gap between the central galaxy and the most massive satellite is smaller for more massive groups/clusters \citep[see, e.g.,][]{Hearin+2013b, Rodriguez-Puebla+2013}. These points indicate that while central/satellite designation is a potential problem for group finders, affecting more massive groups/clusters, it should not significantly affect our clustering measurements of centrals as a function of mass since a misidentified central galaxy will have similar properties to the true central galaxy.

In addition to the central/satellite designation, there are indeed errors when it comes to group membership. As discussed in \citet{Campbell+2015}, group finders may impose fracturing and/or fusing onto their groups. In such cases, members of a single halo may be misidentified as belonging to separate groups (fracturing) or members may be assigned to the same group that do not share the same halo (fusing). In the case of fracturing, there is necessarily a satellite that becomes misidentified as a central. Since satellites tend to be quiescent, we may then expect that this would result in enhanced clustering for quiescent centrals, contributing to the minimal difference we see between the star-forming and quiescent populations at larger scales (this is certainly the case for clustering signals we see at smaller scales for centrals). However, it is difficult to predict the net effect of fracturing and fusing on our results -- and whether it is color dependent -- without performing systematic tests on a mock catalogue, which we leave to future work.

\subsubsection{Group catalogue comparison} \label{section:group_catalog_comparison}
The simplest metric for comparing different group catalogues is to compare the central or satellite designation of a given galaxy across the catalogues. Note that not every galaxy in our sample has a designation in every group catalogue. About 3\percent{} of galaxies have designations in only two catalogues, 3\percent{} have no designations, and fewer than 1\percent{} have designations in only one catalogue. For comparison, we consider only galaxies that have a designation in all three catalogues. We find generally good agreement across the three group catalogues, ranging from $\sim$70\percent{} to 90\percent{} with the agreement improving with mass. The values for each mass bin are shown in Table~\ref{tab:group_catalog_match}.

Another relevant statistic in the group catalogues is the relative numbers of central and satellite galaxies. Group catalogues that identify a higher ratio of centrals to satellites may have a higher rate of contamination where satellites are misidentified as centrals. On the other hand, group catalogues that identify a lower ratio of centrals to satellites may have a more pure sample of centrals. The ratios of centrals to satellites for the three group catalogues are shown in Fig.~\ref{fig:centrals_per_satellite}. All three catalogues are fairly similar, though \citetalias{Tempel+2017} has a clear lower ratio.

As a comparison to Fig.~\ref{fig:centrals_ACFs}, we show the same results of ACFs of centrals using \citetalias{Tempel+2017} in Fig.~\ref{fig:tempel_centrals_ACFs}. We see the same lack of a distinct trend in the central ACFs that we saw using the \citetalias{Yang+2012} catalogue. One notable difference, however, is the lack of a signal at small separations. If, indeed, \citetalias{Tempel+2017} provides a more pure sample of only central galaxies, this would suggest the apparent signal at small separations for low-mass centrals in Fig.~\ref{fig:centrals_ACFs} is most likely the result of satellites misidentified as centrals in the \citetalias{Yang+2012} group catalogue. We note also that this apparent signal occurs primarily in the quiescent sub-sample, and we know the satellite fraction is highest when considering quiescent galaxies. Thus, any systematic bias in the misidentification of satellites as centrals would manifest most strongly in the quiescent galaxy sub-sample. Additionally, we can compare to the ACFs of centrals using \citetalias{Rodriguez_Merchan2020}. From Fig.~\ref{fig:centrals_per_satellite}, we know that \citetalias{Rodriguez_Merchan2020} has a ratio of centrals to satellites that is comparable to \citetalias{Yang+2012}, with a slightly higher ratio in the lowest mass bin and a slightly lower ratio in the next three mass bins. In the results of the \citetalias{Rodriguez_Merchan2020} central ACFs, we saw the strongest small separation signal of the three catalogues for low-mass quiescent (and green valley) centrals, as well as a signal in the next three mass bins similar to \citetalias{Yang+2012}. This result is consistent with our expectations of potential satellite contamination based on the ratios of centrals to satellites.

\begin{figure*}
    \centering
    \includegraphics[width=\linewidth]{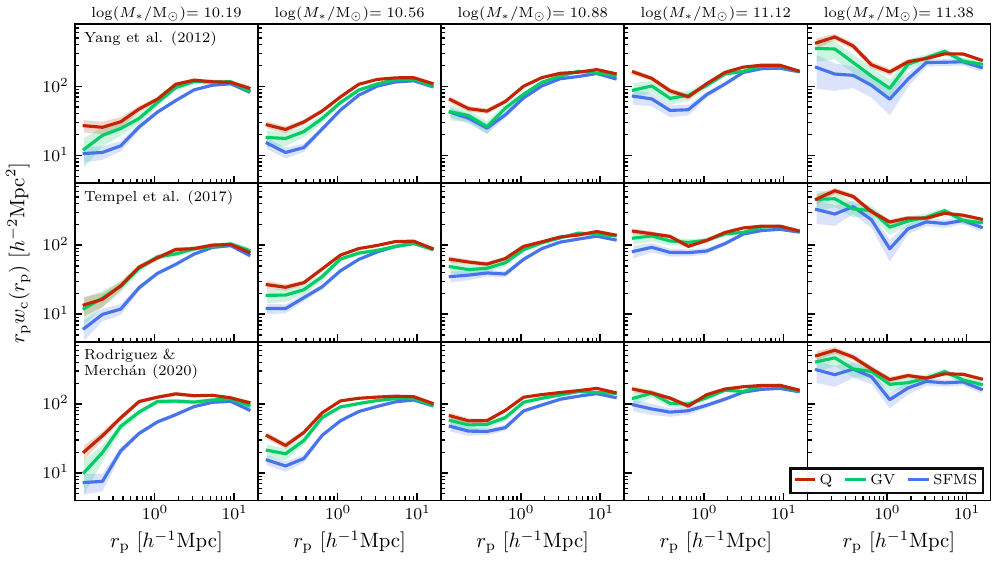}
    \caption{This shows a simplified version of Fig.~\ref{fig:CCFs} where the grid has been collapsed vertically and each row now presents the results of a different group catalogue. Shown are the cross-correlation functions of SFMS (blue), GV (green), and quiescent (red) centrals with all satellites in the same mass bin. This highlights the consistent trend across all group catalogues of increasing clustering at small separations as $\Delta$MS decreases.}
    \label{fig:multi_group_cat_CCFs}
\end{figure*}

In Fig.~\ref{fig:multi_group_cat_CCFs}, we show the results of the CCFs for the three different group catalogues. Each row shows a simplified version of Fig.~\ref{fig:CCFs} for each group catalogue. Since we observe no trends in the CCFs across the SFMS in any of the group catalogues, we show only a single blue line per each grid cell corresponding to the CCF of all SFMS centrals with all satellites. As before, the green lines show the CCFs of GV centrals and the red lines show the CCFs of quiescent centrals. The values of the CCFs are generally consistent across the group catalogues, and, similar to Fig.~\ref{fig:CCFs}, we see the same trend of increasing clustering amplitude in the one-halo term with decreasing $\Delta$MS present in both \citetalias{Tempel+2017} and \citetalias{Rodriguez_Merchan2020}. These results reinforce our conclusion that quiescent central galaxies are more likely to have a greater number of satellites than central galaxies located on the SFMS.

\subsection{Comparison with the literature} 
Projected ACFs of galaxies as a function of sSFR and stellar mass have been measured by \citet[hereafter \citetalias{Coil+2017}]{Coil+2017} and \citet[hereafter \citetalias{Berti+2021}]{Berti+2021}. \citetalias{Coil+2017} study higher redshift galaxies from the PRIMUS and DEEP2 galaxy redshift surveys spanning the redshift range $0.2 < z < 1.2$. \citetalias{Berti+2021} study lower redshift SDSS galaxies in the redshift range $0.02 < z < 0.04$ using the same stellar mass and SFR measurements we use in this paper. While different galaxy samples are studied in these two papers, they ultimately draw similar conclusions about the clustering trends with sSFR and stellar mass: namely, that clustering amplitude is as strong (or stronger) a function of sSFR as it is of stellar mass. However, this conclusion is based on ACFs of \emph{all} galaxies without distinguishing between centrals and satellites. As we have shown in Fig.~\ref{fig:all_ACFs}, we confirm the presence of a strong sSFR-dependent clustering amplitude when considering centrals and satellites together. However, as we have shown in Fig.~\ref{fig:all_ACF_components}, this trend is almost entirely produced by satellites which are predominantly quiescent galaxies. This possible explanation for the observed trend is mentioned in Section~6.3 of \citetalias{Coil+2017}, and we confirm it in this paper for the stellar mass and sSFR ranges we study.

Another conclusion of \citetalias{Coil+2017} and \citetalias{Berti+2021} is the existence of intrasequence relative bias (ISRB), i.e., that within the star-forming (or quiescent) population, galaxies with higher sSFR have lower clustering amplitude than galaxies with lower sSFR. While we do not attempt to make any determination about this for quiescent galaxies, we can compare results of similar sub-samples for star-forming galaxies. In general, we do not observe ISRB within the SFMS, even when measuring clustering using centrals and satellites together. However, we do not believe our results are necessarily at odds with their conclusion. This effect may be more prominent at lower masses as \citetalias{Coil+2017} and \citetalias{Berti+2021} use samples that go well below our lower limit of $\log(M_*/\Msun)=10$. We attempt here to illustrate this but note that the wider mass bins used by \citetalias{Coil+2017} and \citetalias{Berti+2021} make it difficult to give this as the definitive explanation.

We can investigate this by comparing our results with fig.~1 of \citetalias{Berti+2021}. In the rightmost panels (their `$M_*$/sSFR grid'), we can compare the results of the black and blue regions, which are in a fixed mass bin from $9.75 < \log(M_*/\Msun) < 10.4$ with mean sSFRs of $\log(\text{sSFR/yr}^{-1}) = -9.82$ and $-10.33$, respectively. These regions roughly correspond to star-forming galaxies above and below the center of the SFMS in our lowest mass bin: $10.0 < \log(M_*/\Msun) < 10.375$ with mean sSFRs of $\log(\text{sSFR/yr}^{-1}) \approx -9.9$ and $-10.3$, respectively. While these sub-samples do not directly correspond to each other, they are sufficiently similar for the sake of this comparison. We can see their clustering amplitudes are fairly similar, as we also observed in our results. There is a difference in clustering amplitude at larger separations for the higher and lower sSFR sub-samples, but the difference looks to be within the error bars. However, if we compare their purple and magenta regions, which are at a lower mass of $9.25 < \log(M_*/\Msun) < 9.75$ but in the \emph{same} sSFR bins as the black and blue regions, we see a much larger increase in clustering going from the purple (higher sSFR) to the magenta (lower sSFR) sub-sample. This suggests that the ISRB within the SFMS may indeed be much more prominent at lower stellar masses, which we do not probe in this paper.

A final consideration we give to this effect is the potential mass trend within a mass bin. In this paper, we have chosen narrower mass bins to attempt to limit the extent to which the mean stellar mass within a given mass bin can vary. Clustering amplitude will generally increase as stellar mass increases, so it is important to bear this in mind when attempting to isolate clustering trends with sSFR. We do not attempt to quantify the magnitude of the mass effect on clustering amplitude, but we simply note that the wider mass bins used by \citetalias{Coil+2017} and \citetalias{Berti+2021} are more susceptible to intra-bin mass trends. The aforementioned black and blue sub-samples used in \citetalias{Berti+2021} have a difference in mean stellar mass of 0.08~dex while our comparable bins have a difference of only 0.016~dex. This difference may also play a role in explaining the differences between our results and those of \citetalias{Berti+2021}. We do note, however, that the mean mass difference for the purple and magenta sub-samples is only 0.04~dex, which suggests that intra-bin mass trends in this case likely are not the cause of the ISRB observed in \citetalias{Coil+2017} and \citetalias{Berti+2021}.

We note that fig.~12 of \citetalias{Berti+2021} plots the 3D ACFs of central and satellite galaxies separately predicted by the revised \textsc{UniverseMachine}, showing that the ACFs of both centrals and satellites are higher for quiescent and lower for star-forming galaxies. In initial tests, we found that the SFR-dependence of the clustering signal for central galaxies in the \textsc{UniverseMachine} is largely driven by backsplash haloes that have lost at least 10\percent{} of their peak mass and that non-backsplash haloes have nearly no SFR-dependence in their clustering. As backsplash galaxies are expected to be drawn from the same population as satellite galaxies (and hence have much higher quenched fractions than field galaxies), it would be difficult to remedy the clustering of the \textsc{UniverseMachine} by adjusting backsplash galaxies alone. This would suggest, along the same lines as in \citet{Cui+2021}, that non-backsplash centrals must actually have an inverse SFR-dependence in their clustering so that the clustering of the entire central sample has no SFR dependence (see also \citealt{ODonnell+2021, ODonnell+2022}, who reached similar conclusions). We will follow up on this possibility in future work.

A similar analysis to \citetalias{Coil+2017} and \citetalias{Berti+2021} was done by \citet[hereafter \citetalias{Berti+2019}]{Berti+2019}, who use PRIMUS galaxies around $z\sim0.35$ and $z\sim0.7$. \citetalias{Berti+2019} search for trends in clustering with sSFR for isolated primary (IP) galaxies, defined as those with no neighboring galaxies above a stellar mass threshold within a surrounding cylindrical volume. These IPs were used as proxies for central galaxies to test if the trends found in \citetalias{Coil+2017} hold without the inclusion of satellites. \citetalias{Berti+2019} find that quiescent IPs are more clustered than star-forming IPs, with a relative bias of $1.7\pm0.07$ at $z\sim0.35$ and $1.31\pm0.16$ at $z\sim0.7$ (see fig.~6 of \citetalias{Berti+2019}). These biases are stronger than what we find in this paper for central galaxies. While we emphasize that these results are at higher redshifts and evolutionary effects can lead to differences, a possible explanation is that the biases are enhanced by the presence of satellites in their samples, which we discuss below.

As part of the study, \citetalias{Berti+2019} use a mock catalogue to test for satellite contamination in their method of choosing IPs. Fig.~10 of \citetalias{Berti+2019} shows that at $z\sim0.35$ there is a $\sim$20\percent{} contamination for quiescent IPs but only about half as much for star-forming IPs. At $z\sim0.7$, there is less difference in contamination between quiescent and star-forming IPs, though it grows larger for higher stellar masses. If the main driver of the clustering difference between quiescent and star-forming galaxies is satellites, it would be expected that the higher contamination difference in their $z\sim0.35$ sample would create a greater relative bias between quiescent and star-forming IPs than their $z\sim0.7$ sample, which is indeed what \citetalias{Berti+2019} find. Further, under this same assumption, it would be expected to not find significant ISRB, assuming the satellite contamination does not vary much within a given population. This is again what \citetalias{Berti+2019} find, similar to what we find in this paper for star-forming centrals. With this in mind, we believe our results may be consistent with \citetalias{Berti+2019} despite initially seeming to disagree. Additional testing would be required to determine more specifically how much difference in central or IP clustering results from a differing satellite contamination between quiescent and star-forming populations. Preliminary results from our clustering analysis in large-volume cosmological hydrodynamic simulations (such as Illustris TNG100) show that the clustering of star-forming and quiescent central galaxies is consistent with our results at $z\sim0$. We also find little difference in the clustering of star-forming and quiescent centrals in TNG100 at $z\sim0.5$

\section{Conclusions} \label{section:conclusions}
In this paper, we analyzed the clustering properties of galaxies in the sSFR--$M_*$ plane by calculating two-point auto-correlation functions (ACFs) of all, central, and satellite galaxies and cross-correlation functions (CCFs) of centrals with satellites as a function of distance from the mean of the star-forming main sequence (SFMS) in various mass bins. These calculations were done using a spectroscopic sample of galaxies from the SDSS in the redshift range $0.02<z<0.2$ with stellar masses and star formation rates from the MPA-JHU catalogue \citep{Kauffmann+2003, Brinchmann+2004}. To define central and satellite galaxies, we used three different group catalogues from \citet{Yang+2012}, \citet{Tempel+2017}, and \citet{Rodriguez_Merchan2020}. In each of these group catalogues, central galaxies were defined as the galaxy with the largest stellar mass in each group. For details on the catalogues and data selection, see Section~\ref{section:data}. We draw the following conclusions from our analysis based on the \citet{Yang+2012} group catalogue for galaxies with stellar masses in the range $10.0 < \log(M_*/\Msun) < 11.5$:
\begin{enumerate}
    \item At a fixed stellar mass, ACFs of all (central and satellite) galaxies together depend strongly on sSFR (Fig.~\ref{fig:all_ACFs}). Green valley and quiescent galaxies have greater clustering amplitudes than SFMS galaxies, with quiescent clustering amplitudes increasing up to as much as an order of magnitude larger than that of SFMS galaxies.
    
    \item When considering only central galaxies, ACFs show little to no dependence on sSFR at a fixed stellar mass (Figs~\ref{fig:centrals_ACFs} and \ref{fig:tempel_centrals_ACFs}). The ACFs do, however, increase with stellar mass. Since galaxy stellar mass increases monotonically with halo mass, and halo clustering also increases with halo mass (known as halo bias), we conclude that the increased galaxy clustering with stellar mass is driven by halo bias. However, halo clustering can also vary depending on secondary factors such as formation redshift and accretion history (known as assembly bias). Because the observed central galaxy ACFs do not vary with sSFR at a fixed stellar mass, this suggests that assembly bias is not producing any significant net effect in central galaxy clustering in the sSFR--$M_*$ plane.
    
    \item By calculating the ACFs of centrals and satellites separately, we found that satellites are the main contributor to the strong sSFR dependence of all-galaxy ACFs (Fig.~\ref{fig:all_ACF_components}). Since satellites are innately highly clustered and tend to be quiescent, they drive up the clustering amplitudes as lower sSFR samples are considered at a fixed stellar mass.
    
    \item Cross-correlations of central galaxies as a function of sSFR with satellite galaxies of any sSFR show that more quiescent centrals have higher clustering amplitudes in the one-halo term than SFMS centrals in all stellar mass bins studied in this paper (Fig.~\ref{fig:CCFs}). CCFs of quiescent centrals are larger than those of SFMS centrals by up to a factor of $\sim$3 in the one-halo term. This suggests that quiescent centrals tend to host more satellites in their dark matter haloes on average than SFMS centrals. We confirmed this conclusion by directly calculating $\langle N_\text{sat}\rangle$ as a function of sSFR (Fig.~\ref{fig:Nsat}). Within the context of the halo occupation distribution, this implies that quiescent centrals reside in higher mass haloes than SFMS centrals of the same stellar mass, which is consistent with observations.
\end{enumerate}

All results were additionally checked using the group catalogues from \citet{Tempel+2017} and \citet{Rodriguez_Merchan2020} to test for robustness. We find that our conclusions are valid regardless of the choice in group catalogue employed for clustering analysis (see Section~\ref{section:group_catalog_comparison} and Fig.~\ref{fig:multi_group_cat_CCFs}).

We used these conclusions to test whether different models of the galaxy--halo connection could reproduce, at the same time, the sSFR--$M_*$ plane and the ACFs, CCFs, and $\langle N_\text{sat}\rangle$ as a function of distance from the mean SFMS. Additionally, we directly calculated the resultant stellar-to-halo mass relations (SHMRs) from these models for star-forming and quiescent galaxies. We assigned sSFR values to dark matter haloes in the Bolshoi--Planck $N$-body simulation using three different models based on (1) halo accretion rate, $\dot{M}_\text{h}$, (2) halo concentration, $C_\text{vir}$, and (3) peak circular velocity, $V_\text{peak}$. From comparing the models to the observations, we draw the following conclusions:
\begin{enumerate}
    \item ACFs of central galaxies from the three different models of sSFR all show \emph{similar} results that agree well with observations (Fig.~\ref{fig:galhalo_ACFs}). It may be expected that the assembly bias built into the $\dot{M}_\text{h}$ and $C_\text{vir}$ models would lead to trends in the ACFs with sSFR at a fixed stellar mass, but no trends were observed. This may be the result of competing biases within the models or the assembly bias effect being lessened or removed when binning by \emph{stellar} mass rather than \emph{halo} mass.
    
    \item Modeling sSFR based on $\dot{M}_\text{h}$ or $C_\text{vir}$ did not reproduce trends in the CCFs of observed SDSS galaxies (Fig.~\ref{fig:galhalo_CCFs}). Instead, these models predicted a trend \emph{opposite} to what is observed. We also saw an opposite trend when considering $\langle N_\text{sat}\rangle$ for star-forming and quiescent galaxies (Fig.~\ref{fig:galhalo_Nsat}). However, modeling sSFR based on $V_\text{peak}$ produced CCFs that were consistent with observations and showed quiescent centrals with greater $\langle N_\text{sat}\rangle$ than star-forming centrals at all stellar masses, also agreeing with SDSS observations. These results suggest that the cross-correlation clustering of centrals with satellites in the sSFR--$M_*$ plane may be primarily driven by halo mass, as $V_\text{peak}$ is strongly correlated with halo mass.

    \item While ACFs showed similar results for all models, the capability of CCFs and $\langle N_\text{sat}\rangle$ to differentiate models of sSFR based on halo properties has proven to be a powerful tool for constraining different models of the galaxy--halo connection.

    \item Comparing the SHMRs of the different models showed that all models agree with observations of the SHMR of quiescent galaxies, but only the $V_\text{peak}$ model agreed that star-forming galaxies reside in haloes of lower mass than quiescent galaxies at a fixed stellar mass. The accretion rate and concentration models both predicted that star-forming galaxies reside in haloes of \emph{higher} mass at a fixed stellar mass, again, opposite of what the observations show.

    \item In the $V_\text{peak}$ model, haloes hosting SFMS and quiescent galaxies show minimal difference in formation redshift. This suggests that the $V_\text{peak}$ model is primarily associated with halo mass and rules out being influenced by assembly bias related to the relationship between $V_\text{peak}$ and $C_\text{vir}$.
\end{enumerate}

In our $V_\text{peak}$ model, the difference in clustering for quiescent and star-forming centrals is driven primarily by a segregation in the SHMR, wherein quiescent centrals reside in haloes of higher mass than star-forming centrals at the same stellar mass. The evidence obtained from both observations and our models of sSFR suggests that the $V_\text{peak}$ model offers a realistic model of how massive ($\log(M_*/M_\odot)\geq 10)$ centrals inhabit dark matter haloes. 

Early-forming haloes form in denser regions of the Universe, and they are therefore more clustered. It is tempting to imagine that this assembly bias of early-forming haloes helps explain why galaxies with lower sSFR are more clustered. In more detail, the idea is that early-forming haloes host early-forming galaxies, which also quench early. But if this were true of central galaxies, the quenched centrals would be more correlated than SFMS centrals, which we have shown is not true. This implies that central galaxies in early-forming haloes do not, in general, quench earlier than central galaxies in later-forming haloes of the same mass.

Our finding that the central galaxy ACFs are unchanged as a function of sSFR at a fixed stellar mass leaves the door open for multiple possible explanations that require further testing. While the $V_\text{peak}$ model does not include assembly bias, one could imagine a model that uses both a segregation in the SHMR and assembly bias to reproduce the observed ACFs. With a segregation in the SHMR such that quiescent galaxies reside in more massive haloes than star-forming galaxies at a fixed stellar mass, it may be expected that quiescent centrals should be more clustered than star-forming centrals of the same stellar mass, as a result of halo bias. Since we do not observe this, it suggests, perhaps, that there could be an additional effect increasing the clustering amplitude of star-forming galaxies, leading to an overall lack of clustering trend with sSFR. Interestingly, this would be an example of assembly bias favoring higher clustering for \emph{star-forming} galaxies rather than quiescent galaxies, as is often considered.

An astrophysical example of this is provided by the Simba large-volume cosmological hydrodynamic simulation \citep{SIMBA2019, Cui+2021}, in which late-quenching galaxies live in early-formed haloes and early-quenching galaxies live in later-formed haloes. In Simba, this occurs because of a combination of astrophysical effects: central galaxies in early-forming haloes accrete more cold gas and AGNs in such galaxies are less effective in quenching them, while late-forming haloes have mainly hot-mode accretion and the associated jet-mode AGN feedback quenches them early. For example, Section~3 of the Supplementary Information for \citet{Cui+2021} describes three types of haloes of the same mass at $z=0$, in which the central galaxies in the early-forming haloes quench last, and the central galaxies in the late-forming haloes quench earliest. Consistent with this, \citet{WangChenLiYang2023} note that both observations and the EAGLE simulation indicate that early-formed haloes tend to have a higher ratio of stellar mass to halo mass, and their analysis of SDSS galaxies implies that late-formed haloes tend to host quiescent galaxies. We leave to a future paper further study of such large-volume simulations, including the ACFs and CCFs of their central and satellite galaxies. Our preliminary finding is that the ACFs of star-forming and quiescent central galaxies in the Simba, EAGLE, and IllustrisTNG simulations are consistent with our analysis of the SDSS.

\section*{Acknowledgements}
We are grateful for helpful conversations with Doug Hellinger, Kai Wang, and Joanna Woo. We also thank the referee, Alison Coil, for helpful comments that improved this paper. ARP and VAR acknowledge financial support from CONACyT ``Ciencia Basica'' grant 285721, CONAHCyT ``Ciencia de Frontera'' grant G-543, and DGAPA-PAPIIT IN106823, IN106124 and IN106924. We thank contributors to the \textsc{Python} programming language\footnote{\url{https://www.python.org/}}, \textsc{SciPy}\footnote{\url{https://www.scipy.org/} \citep{2020SciPy-NMeth}}, \textsc{NumPy}\footnote{\url{https://numpy.org/} \citep{2020NumPy-Array}}, \textsc{Matplotlib}\footnote{\url{https://matplotlib.org/} \citep{Hunter:2007}}, and the free and open-source community.

\section*{Data Availability}
The MPA-JHU catalogue with stellar masses and SFRs is available at \url{http://www.mpa-garching.mpg.de/SDSS/DR7/}. Additional galaxy data can be downloaded from the SDSS Catalog Archive Server at \url{https://skyserver.sdss.org/CasJobs/}. The group catalogues for \citetalias{Yang+2012}, \citetalias{Tempel+2017}, and \citetalias{Rodriguez_Merchan2020} can be found within their respective papers and references therein. Snapshots of the Bolshoi--Planck simulation at various time steps are available at \url{https://halos.as.arizona.edu/simulations/BolshoiP/hlists/}.

%%%%%%%%%%%%%%%%%%%%%%%%%%%%%%%%%%%%%%%%%%%%%%%%%%

%%%%%%%%%%%%%%%%%%% REFERENCES %%%%%%%%%%%%%%%%%%%
\bibliographystyle{mnras}
\bibliography{main}

%%%%%%%%%%%%%%%%%%%%%%%%%%%%%%%%%%%%%%%%%%%%%%%%%%

%%%%%%%%%%%%%%%%%%% APPENDICES %%%%%%%%%%%%%%%%%%%

\appendix
\section{Sub-volume Completeness}
\begin{figure*}
    \centering
    \includegraphics[width=\linewidth]{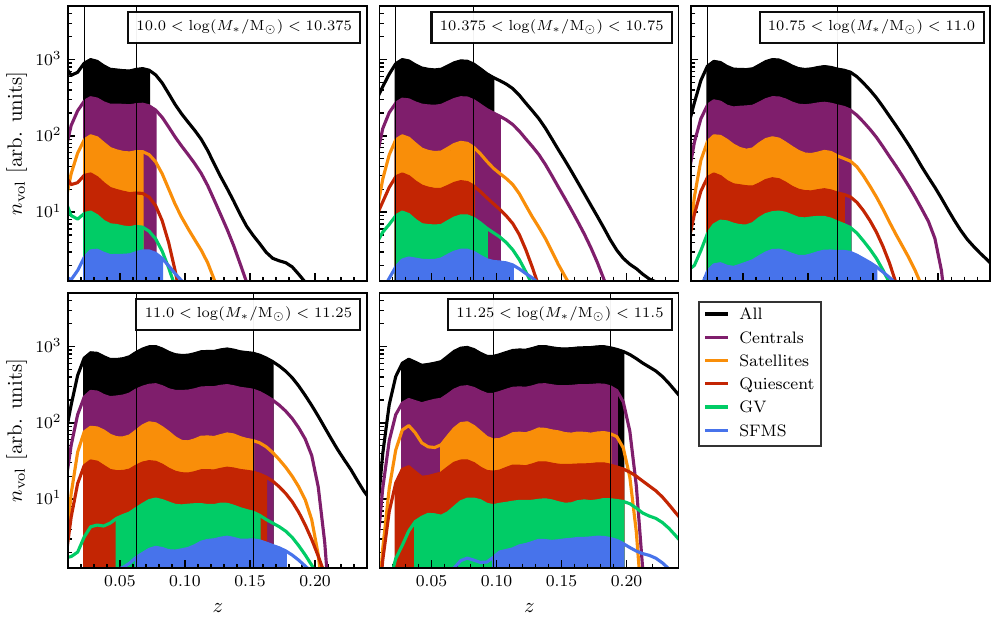}
    \caption{Joint completeness of different sub-samples within each of our five sub-volumes. The full ranges of number densities are shown by the colored lines, and the shaded regions show the limits of completeness for each sub-sample. The vertical black lines denote the chosen redshift limits for each sub-volume, which are selected as the highest low-redshift limits and lowest high-redshift limits that allow for completeness in every sub-sample within each sub-volume. Note that the distributions have been normalized and scaled by increments of 0.5~dex for clarity.}
    \label{fig:completeness_subsamples}
\end{figure*}
\begin{table*}
    \centering
    \caption{List of sub-samples in the sSFR--$M_*$ plane with the corresponding mean and median values of $\log(M_*/\Msun)$ and $\Delta$MS and counts of central and satellite galaxies from \citetalias{Yang+2012}. Note that $N_\text{cen}$ and $N_\text{sat}$ are repeated in the table for ease of comparison of their relative statistics in a given bin.}
    \begin{tabular}{@{\extracolsep{2pt}}c l r r r r r r@{}}
        \hline
        \multirow{2}{*}{$M_*$ Limits [$\log(M_*/\Msun)$]} & \multirow{2}{*}{Sub-sample} & \multicolumn{2}{c}{$\log(M_*/\Msun)$} & \multicolumn{2}{c}{$\Delta$MS [dex]} & \multirow{2}{*}{$N_\text{cen}$} & \multirow{2}{*}{$N_\text{sat}$}\\
        \cline{3-4}\cline{5-6}
        \rule{0pt}{8pt} && Mean & Median & Mean & Median &&\\
        \hline
        \multicolumn{8}{c}{Centrals}\\
        \hline
        \rule{0pt}{8pt}10.0 -- 10.375 & HSF & 10.179 & 10.177 & 0.389 & 0.353 & 1,456 & 437\\
        & UMS & 10.172 & 10.165 & 0.115 & 0.111 & 2,930 & 871\\
        & LMS & 10.181 & 10.174 & -0.120 & -0.117 & 2,989 & 971\\
        & BMS & 10.194 & 10.198 & -0.344 & -0.342 & 1,483 & 607\\
        & GV & 10.194 & 10.192 & -0.705 & -0.691 & 2,111 & 1,367\\
        & Q & 10.210 & 10.220 & -1.404 & -1.406 & 3,729 & 3,855\\
        
        10.375 -- 10.75 & HSF & 10.544 & 10.539 & 0.428 & 0.392 & 2,814 & 761\\
        & UMS & 10.536 & 10.523 & 0.115 & 0.111 & 4,137 & 1,080\\
        & LMS & 10.538 & 10.529 & -0.121 & -0.121 & 4,266 & 1,206\\
        & BMS & 10.548 & 10.540 & -0.348 & -0.347 & 2,380 & 813\\
        & GV & 10.557 & 10.554 & -0.720 & -0.714 & 4,597 & 2,192\\
        & Q & 10.568 & 10.572 & -1.455 & -1.478 & 13,273 & 8,433\\
        
        10.75 -- 11.0 & HSF & 10.859 & 10.852 & 0.474 & 0.425 & 3,614 & 715\\
        & UMS & 10.855 & 10.845 & 0.124 & 0.125 & 3,452 & 724\\
        & LMS & 10.860 & 10.852 & -0.125 & -0.124 & 3,419 & 776\\
        & BMS & 10.864 & 10.857 & -0.349 & -0.348 & 2,546 & 626\\
        & GV & 10.869 & 10.866 & -0.735 & -0.737 & 6,901 & 2,200\\
        & Q & 10.870 & 10.868 & -1.412 & -1.424 & 21,203 & 8,409\\
        
        11.0 -- 11.25 & HSF & 11.096 & 11.084 & 0.503 & 0.454 & 2,703 & 384\\
        & UMS & 11.095 & 11.083 & 0.120 & 0.119 & 2,250 & 357\\
        & LMS & 11.100 & 11.091 & -0.125 & -0.123 & 2,489 & 443\\
        & BMS & 11.104 & 11.095 & -0.353 & -0.351 & 2,202 & 471\\
        & GV & 11.108 & 11.101 & -0.736 & -0.739 & 6,953 & 1,521\\
        & Q & 11.112 & 11.105 & -1.353 & -1.360 & 23,229 & 5,903\\
        
        11.25 -- 11.5 & HSF & 11.334 & 11.319 & 0.534 & 0.475 & 954 & 115\\
        & UMS & 11.342 & 11.326 & 0.117 & 0.119 & 815 & 96\\
        & LMS & 11.341 & 11.326 & -0.131 & -0.132 & 1,066 & 114\\
        & BMS & 11.345 & 11.334 & -0.351 & -0.350 & 1,123 & 144\\
        & GV & 11.349 & 11.336 & -0.768 & -0.796 & 3,637 & 482\\
        & Q & 11.350 & 11.339 & -1.294 & -1.296 & 15,007 & 2,135\\
        \hline
        \multicolumn{8}{c}{Satellites}\\
        \hline
        \rule{0pt}{8pt}10.0 -- 10.375 & HSF & 10.172 & 10.168 & 0.404 & 0.358 & 1,456 & 437\\
        & UMS & 10.167 & 10.165 & 0.112 & 0.109 & 2,930 & 871\\
        & LMS & 10.170 & 10.163 & -0.126 & -0.130 & 2,989 & 971\\
        & BMS & 10.191 & 10.195 & -0.349 & -0.345 & 1,483 & 607\\
        & GV & 10.179 & 10.176 & -0.723 & -0.722 & 2,111 & 1,367\\
        & Q & 10.191 & 10.191 & -1.452 & -1.467 & 3,729 & 3,855\\
        
        10.375 -- 10.75 & HSF & 10.539 & 10.524 & 0.433 & 0.403 & 2,814 & 761\\
        & UMS & 10.537 & 10.526 & 0.115 & 0.113 & 4,137 & 1,080\\
        & LMS & 10.533 & 10.516 & -0.123 & -0.119 & 4,266 & 1,206\\
        & BMS & 10.543 & 10.535 & -0.352 & -0.354 & 2,380 & 813\\
        & GV & 10.547 & 10.539 & -0.732 & -0.731 & 4,597 & 2,192\\
        & Q & 10.555 & 10.551 & -1.499 & -1.527 & 13,273 & 8,433\\
        
        10.75 -- 11.0 & HSF & 10.859 & 10.849 & 0.477 & 0.428 & 3,614 & 715\\
        & UMS & 10.855 & 10.849 & 0.119 & 0.119 & 3,452 & 724\\
        & LMS & 10.856 & 10.851 & -0.125 & -0.125 & 3,419 & 776\\
        & BMS & 10.863 & 10.857 & -0.356 & -0.357 & 2,546 & 626\\
        & GV & 10.866 & 10.860 & -0.737 & -0.744 & 6,901 & 2,200\\
        & Q & 10.863 & 10.858 & -1.437 & -1.453 & 21,203 & 8,409\\
        
        11.0 -- 11.25 & HSF & 11.088 & 11.073 & 0.499 & 0.458 & 2,703 & 384\\
        & UMS & 11.091 & 11.078 & 0.109 & 0.102 & 2,250 & 357\\
        & LMS & 11.098 & 11.086 & -0.130 & -0.126 & 2,489 & 443\\
        & BMS & 11.101 & 11.090 & -0.354 & -0.354 & 2,202 & 471\\
        & GV & 11.100 & 11.089 & -0.728 & -0.722 & 6,953 & 1,521\\
        & Q & 11.101 & 11.091 & -1.379 & -1.385 & 23,229 & 5,903\\
        
        11.25 -- 11.5 & HSF & 11.319 & 11.307 & 0.516 & 0.456 & 954 & 115\\
        & UMS & 11.327 & 11.304 & 0.118 & 0.124 & 815 & 96\\
        & LMS & 11.338 & 11.321 & -0.129 & -0.127 & 1,066 & 114\\
        & BMS & 11.339 & 11.325 & -0.354 & -0.358 & 1,123 & 144\\
        & GV & 11.333 & 11.320 & -0.765 & -0.785 & 3,637 & 482\\
        & Q & 11.338 & 11.325 & -1.308 & -1.313 & 15,007 & 2,135\\
        \hline
    \end{tabular}
    \label{tab:sample_statistics}
\end{table*}

The main goal of this paper was to study the clustering properties of centrals in the sSFR--$M_*$ plane. Doing this involved breaking down our full galaxy sample into different sub-samples: primarily central, satellite, star-forming, green valley, and quiescent galaxies. Our objective was to isolate clustering trends in sSFR and therefore worked within five bins of stellar mass. Each bin would then represent a volume-limited sample complete in stellar mass over the mass range of the bin. To create such sub-volumes, we used the number density distributions as a function of redshift, $n_\text{vol}(z)$, for each of the various sub-samples to find redshift bounds within which every sub-sample was complete. The specifics for our definition of completeness are described in Section~\ref{section:data_selection}. The $n_\text{vol}(z)$ distributions used to define the redshift limits of our sub-volumes are shown in Fig.~\ref{fig:completeness_subsamples} (note that they have been normalized and scaled for clarity). The colored lines show $n_\text{vol}(z)$ for the full redshift range, and the shaded regions show the limits where each sub-sample is complete. The final limits of each sub-volume, shown as vertical black lines, are taken as the redshift limits within which every sub-sample is complete. Statistics for each sub-volume are shown in Table~\ref{tab:sample_statistics} for central and satellite galaxies.

\section{Central galaxy auto-correlation relative bias}
In Table~\ref{tab:bias}, we provide values of the two-halo term (averaged over $1 < r_\text{p} / h^{-1}\text{Mpc} < 10$) relative bias of central galaxy ACFs of the \citetalias{Yang+2012}, \citetalias{Tempel+2017}, \citetalias{Rodriguez_Merchan2020} group catalogues and all galaxies (including satellites). 

\begin{table*}
    \centering
    \caption{Relative bias averaged over $1 < r_\text{p}/h^{-1}\text{Mpc} < 10$ of different sub-samples for central galaxies of the \citetalias{Yang+2012}, \citetalias{Tempel+2017}, \citetalias{Rodriguez_Merchan2020} group catalogues and all galaxies (including satellites). The relative bias of a sub-sample is defined with respect to the clustering of the entire mass bin the sub-sample is in (see equation~(\ref{eq:bias}))}
    \begin{tabular}{c l c c c c}
        \hline
        \rule{0pt}{8pt}$\log(M_*/\Msun)$ & Sub-sample & Y12 & T17 & R\&M20 & All Galaxies\\
        \hline
        \rule{0pt}{8pt}10.0--10.375  & SFMS & $0.97\pm0.02$ & $0.95\pm0.02$ & $0.90\pm0.01$ & $0.79\pm0.01$\\
        \rule{0pt}{8pt}              & HSF  & $0.91\pm0.09$ & $0.85\pm0.12$ & $0.81\pm0.07$ & $0.73\pm0.03$\\
        \rule{0pt}{8pt}              & UMS  & $1.01\pm0.05$ & $0.98\pm0.06$ & $0.89\pm0.04$ & $0.83\pm0.02$\\
        \rule{0pt}{8pt}              & LMS  & $0.98\pm0.05$ & $0.98\pm0.06$ & $0.92\pm0.03$ & $0.83\pm0.02$\\
        \rule{0pt}{8pt}              & BMS  & $0.78\pm0.08$ & $0.81\pm0.11$ & $0.86\pm0.06$ & $0.80\pm0.03$\\
        \rule{0pt}{8pt}              & GV   & $1.05\pm0.06$ & $1.10\pm0.09$ & $1.13\pm0.05$ & $1.05\pm0.02$\\
        \rule{0pt}{8pt}              & Q    & $1.08\pm0.04$ & $1.06\pm0.05$ & $1.16\pm0.04$ & $1.31\pm0.03$\\
        \rule{0pt}{8pt}10.375--10.75 & SFMS & $0.96\pm0.02$ & $0.95\pm0.02$ & $0.91\pm0.01$ & $0.80\pm0.01$\\
        \rule{0pt}{8pt}              & HSF  & $0.99\pm0.06$ & $1.03\pm0.08$ & $0.92\pm0.06$ & $0.83\pm0.02$\\
        \rule{0pt}{8pt}              & UMS  & $0.93\pm0.04$ & $0.88\pm0.05$ & $0.91\pm0.04$ & $0.79\pm0.02$\\
        \rule{0pt}{8pt}              & LMS  & $0.95\pm0.04$ & $0.97\pm0.06$ & $0.89\pm0.04$ & $0.81\pm0.02$\\
        \rule{0pt}{8pt}              & BMS  & $0.95\pm0.07$ & $0.92\pm0.09$ & $0.97\pm0.07$ & $0.87\pm0.03$\\
        \rule{0pt}{8pt}              & GV   & $0.96\pm0.04$ & $0.99\pm0.05$ & $1.01\pm0.04$ & $0.97\pm0.02$\\
        \rule{0pt}{8pt}              & Q    & $1.08\pm0.02$ & $1.08\pm0.02$ & $1.12\pm0.02$ & $1.17\pm0.01$\\
        \rule{0pt}{8pt}10.75--11.0   & SFMS & $0.91\pm0.02$ & $0.93\pm0.02$ & $0.90\pm0.02$ & $0.84\pm0.01$\\
        \rule{0pt}{8pt}              & HSF  & $0.87\pm0.06$ & $0.93\pm0.07$ & $0.89\pm0.06$ & $0.81\pm0.03$\\
        \rule{0pt}{8pt}              & UMS  & $0.94\pm0.06$ & $0.92\pm0.08$ & $0.90\pm0.07$ & $0.83\pm0.03$\\
        \rule{0pt}{8pt}              & LMS  & $0.95\pm0.06$ & $1.00\pm0.08$ & $0.92\pm0.07$ & $0.87\pm0.03$\\
        \rule{0pt}{8pt}              & BMS  & $0.89\pm0.08$ & $0.91\pm0.10$ & $0.87\pm0.08$ & $0.91\pm0.04$\\
        \rule{0pt}{8pt}              & GV   & $0.99\pm0.03$ & $1.00\pm0.04$ & $1.01\pm0.03$ & $0.99\pm0.02$\\
        \rule{0pt}{8pt}              & Q    & $1.06\pm0.01$ & $1.05\pm0.02$ & $1.07\pm0.01$ & $1.09\pm0.01$\\
        \rule{0pt}{8pt}11.0--11.25   & SFMS & $0.90\pm0.02$ & $0.90\pm0.03$ & $0.90\pm0.02$ & $0.86\pm0.01$\\
        \rule{0pt}{8pt}              & HSF  & $0.85\pm0.08$ & $0.89\pm0.09$ & $0.86\pm0.07$ & $0.75\pm0.05$\\
        \rule{0pt}{8pt}              & UMS  & $0.83\pm0.11$ & $0.91\pm0.11$ & $0.90\pm0.10$ & $0.86\pm0.06$\\
        \rule{0pt}{8pt}              & LMS  & $0.81\pm0.09$ & $0.83\pm0.09$ & $0.82\pm0.08$ & $0.79\pm0.05$\\
        \rule{0pt}{8pt}              & BMS  & $0.89\pm0.10$ & $1.00\pm0.11$ & $0.99\pm0.10$ & $0.91\pm0.05$\\
        \rule{0pt}{8pt}              & GV   & $0.97\pm0.03$ & $0.99\pm0.04$ & $0.97\pm0.04$ & $0.98\pm0.02$\\
        \rule{0pt}{8pt}              & Q    & $1.05\pm0.01$ & $1.05\pm0.01$ & $1.05\pm0.01$ & $1.06\pm0.01$\\
        \rule{0pt}{8pt}11.25--11.5   & SFMS & $0.84\pm0.05$ & $0.86\pm0.05$ & $0.83\pm0.05$ & $0.80\pm0.03$\\
        \rule{0pt}{8pt}              & HSF  & $-$           & $-$           & $-$           & $-$          \\
        \rule{0pt}{8pt}              & UMS  & $-$           & $-$           & $-$           & $-$          \\
        \rule{0pt}{8pt}              & LMS  & $1.02\pm0.22$ & $1.26\pm0.27$ & $1.05\pm0.25$ & $1.10\pm0.18$\\
        \rule{0pt}{8pt}              & BMS  & $1.05\pm0.24$ & $1.12\pm0.25$ & $1.08\pm0.24$ & $0.95\pm0.14$\\
        \rule{0pt}{8pt}              & GV   & $0.99\pm0.06$ & $1.05\pm0.06$ & $0.99\pm0.06$ & $0.99\pm0.05$\\
        \rule{0pt}{8pt}              & Q    & $1.06\pm0.02$ & $1.06\pm0.02$ & $1.06\pm0.02$ & $1.05\pm0.02$\\     
        \hline
    \end{tabular}
    \label{tab:bias}
\end{table*}

%%%%%%%%%%%%%%%%%%%%%%%%%%%%%%%%%%%%%%%%%%%%%%%%%%

% Don't change these lines
\bsp	% typesetting comment
\label{lastpage}
\end{document}